\documentclass[12pt]{article}

\usepackage{amssymb}
\usepackage{bm}
\usepackage[usenames]{color}
\usepackage{arydshln}
\usepackage{amsmath}
\usepackage{color}
\usepackage{graphics}
\usepackage{amsthm}
\usepackage{ascmac}
\usepackage{amsmath}
\usepackage{amsfonts}
\usepackage{amssymb}
\usepackage{multicol}
\usepackage{epsfig}
\usepackage{here}
\usepackage{lscape}
\usepackage{authblk}

\usepackage{setspace}
\usepackage{listings}
\usepackage{ulem}
\makeatletter

\@addtoreset{equation}{section}
\makeatother
\newtheorem{defi}{Definition}
\newtheorem{theo}{Theorem}

\newtheorem{lemm}{Lemma}
\newtheorem{prop}{Proposition}

\newcommand{\indep}{\mathop{\perp\!\!\!\perp}}
\newcommand{\notindep}{\mathop{\not\!\perp\!\!\!\perp}}
\newcommand{\bld}{\boldsymbol}

\usepackage[top=1in,bottom=1in,left=1in,right=1in]{geometry}

\title{Valid Instrumental Variables Selection Methods using Negative Control Outcomes and Constructing Efficient Estimator}
\author[1,2]{Shunichiro Orihara}

\affil[1]{Graduate School of Data Science, Yokohama City University, Kanagawa, Japan}
\affil[2]{Department of Health Data Science, Tokyo Medical University, Tokyo, Japan}

\date{}

\begin{document}
\allowdisplaybreaks[4]
\begin{singlespace}
\maketitle
\end{singlespace}

\section*{Abstract}
In observational studies, instrumental variable (IV) methods are commonly applied when there exists some unmeasured covariates. In Mendelian Randomization (MR), constructing an allele score by using many single nucleotide polymorphisms (SNPs) is often implemented; however, there are risks estimating biased causal effects by including some invalid IVs. Invalid IVs are candidates of IVs associated with some unobserved variables. To solve this problem, we propose a novel strategy in this paper: using Negative Control Outcomes (NCOs) as auxiliary variables. By using NCOs, we can essentialy select only valid IVs and exclude invalid IVs without any information of IV candidates. We also propose the new two-step estimating procedure and prove the semiparametric efficiency. We demonstrate the superior performance of the proposed estimator compared with existing estimators via simulation studies.

\vspace{0.5cm}
\noindent
{\bf Keywords}: Exclusion restriction, Instrumental variable, Mendelian randomization, Negative control outcome, Semiparametric efficiency, Variable selection, Unmeasured covariates

\section{Introduction}
In observational studies, we sometimes suffer from the situations where some covariates are not observed; i.e., there exists some unmeasured covariates (or confounders; hereafter, we call ``covariates"). In this situation, IV methods are commonly applied. Regarding IV methods for applying to biometrics and related fields, some theoretical results and applications have been appeared in recent years (Brookhart and Schneeweiss, 2007, Baiocchi et al., 2014, Kang et al., 2016, Bowden et al., 2016 Burgess et al., 2017, Hartwig et al., 2017, and Orihara et al., 2022). Figure \ref{fig1} shows the situation discussed in IV contexts. It also explains the three conditions that must be met by the IV.

\begin{figure}[h]
\begin{center}
\begin{tabular}{c}
\includegraphics[width=14cm]{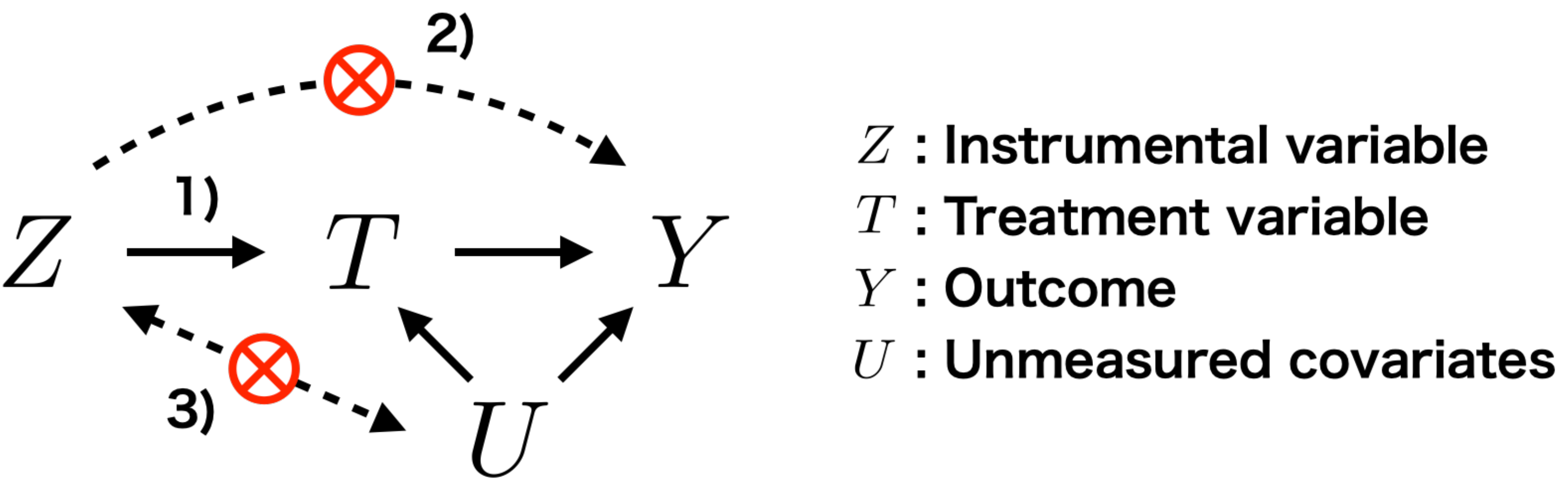}
\end{tabular}
\label{fig1}
\caption{The variable relationship discussed in IV contexts}
\begin{description}
{\footnotesize
\item{\bf IV condition 1):} IVs are related to the treatment
\item{\bf IV condition 2):} IVs are related to the outcome only through the treatment (Exclusion restriction)
\item{\bf IV condition 3):} IVs are independent of umeasured covariates
}
\end{description}
\end{center}
\end{figure}
\noindent
Variables satisfying the above three conditions are called as ``valid IVs", and associated with some unobserved variables are called as ``invalid IVs" (i.e., deviate from IV condition 2) or 3)) in this paper. When using invalid IVs, the estimated causal effects may have some bias. Thus, selecting and using valid IVs are especially important in IV methods.

MR is one of the IV methods used in biometrics and related fields; commonly using SNPs as IVs. Since there are only weak correlations between treatments and each SNP (i.e., weak IV), constructing an allele score by using many SNPs as single IV is one of the solutions (Pierce et al., 2011 and Burgess et al., 2017). An allele score is commonly constructed as a weighted linear combination of SNPs; however, IV estimators may have some bias when some invalid IVs are included in the allele score. It is important to identify invalid IVs and to exclude them before constructing an allele score.

IV condition 1) in Figure \ref{fig1} can be assessed using observed data; whereas, IV condition 2) and 3) are commonly cannot (Baiocchi et al., 2014). The violation of condition 2) is derived from horizontal pleiotropy (Davies et al., 2017 and Sanderson et al., 2022) that is genetic variant related to two or more phenotypes. Therefore, it is possible that the genetic variant is related to outcomes through the unmeasured phenotypes (see Figure 2; the upper dashed line box). To solve the problem, Kang et al., 2016, Bowden et al., 2016, and Hartwig et al., 2017 propose estimators that work well only when there were more than half of valid IVs in candidates of IVs (i.e., the set of candidates of IVs includes valid IVs and invalid IVs).

The violation of IV condition 3) is typically derived from unaware population structure derived from genetic ancestry (Davies et al., 2017). The structure may cause of unmeasured covariates (see Figure 2; the lower dashed line box). DiTraglia, 2016 and Liao, 2013 propose estimators to overcome the problem, however, their methods need to know exactly the set of partial valid IVs (not all valid IVs). Additionally, the method of DiTraglia, 2016 cannot select the valid IVs exactly since the information criterion selects variables from candidates of IVs so that the mean squared error of an estimator of causal effects becomes small.
\begin{figure}[h]
\begin{center}
\begin{tabular}{c}
\includegraphics[width=14cm]{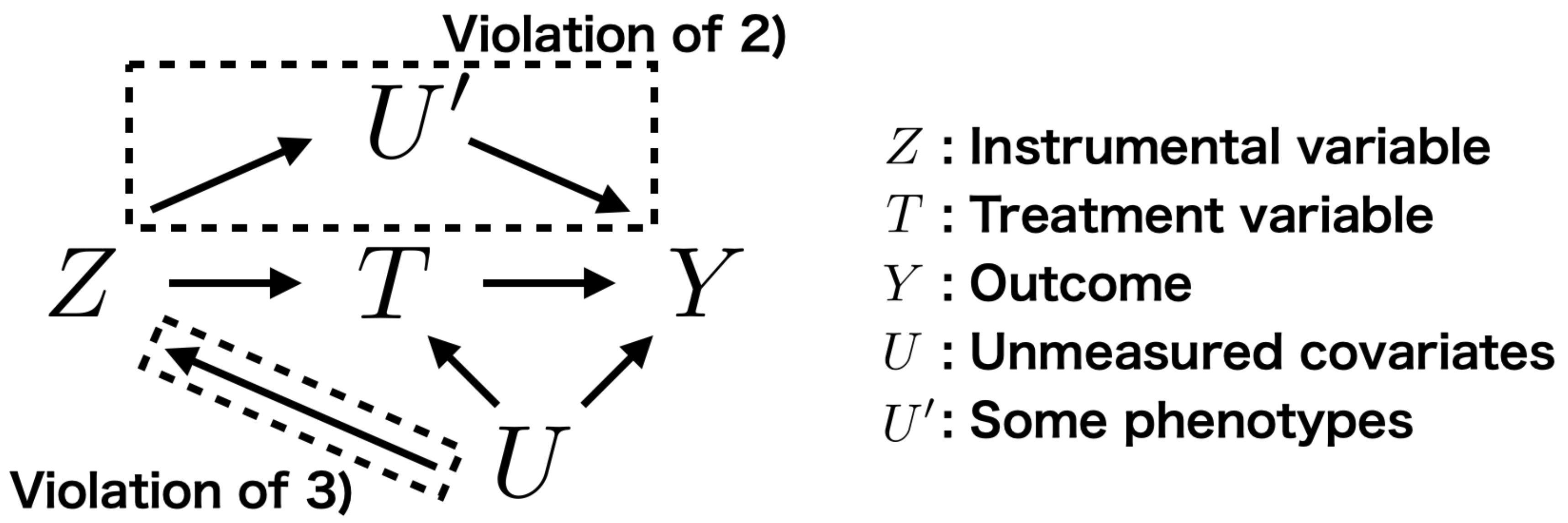}
\end{tabular}
\label{fig2}
\caption{Relationship of variables under some violations}
\end{center}
\end{figure}

The two violations are usually discussed dividedly; however, the mechanism of violations is somewhat similar. In other words, the violations are derived from the non-zero associations between invalid IVs and unobserved variables: unmeasured phenotypes or unmeasured covariates. Therefore, we consider that the two violations should not be discussed separately. In this paper, we propose a novel strategy in this paper: using NCO (e.g., Tchetgen Tchetgen, 2014) as auxiliary variables. By using NCOs, the covariances between IVs and unmeasured phenotypes or unmeasured covariates can be detected all at once. In this paper, we propose the new two-step estimating procedure: an allele score is estimated in the first step, and an outcome model is estimated in the second step. We also prove that our proposed estimator has the same asymptotic variance as Generalized Method of Moments (GMM, see Imbens, 2000); i.e., the semiparametric efficiency. Importantly, any information of IV candidates need not to be detected in advance; our methods are more useful than the previous methods when using NCO.

The remainder of the paper proceeds as follows. In section 2, we discuss the situation where all candidates of IVs are valid. We show that parameters of a linear combination of IVs can be estimated through a linear estimating equation, and our proposed method has the same asymptotic variance as GMM. In section 3, we explain a selection method of the valid IVs without prior informations, and the method retains useful properties. In section 4, we confirm properties of our method and previous methods through simulation. Supplemental information of the main manuscript are given in appendix.

\section{Situations where all instrumental variables are valid}
At first, we discuss the situation where all candidates of IVs are valid; we need not select valid IVs. Let $n$ be the sample size. $T\in\mathcal{T}\subset\mathbb{R}$, $\bld{X}\in\mathbb{R}^{p}$, $U\in\mathbb{R}$, $\bld{Z}\in\mathbb{R}^{K}$, and $Y\in\mathcal{Y}\subset\mathbb{R}$ denote the treatment, a vector of covariates, an unobserved variable, a vector of IVs, and an observed outcome respectively, where the r.v.s have appropriate moment conditions. Note that when $\mathcal{T}=\{0,\, 1\}$, it means binary treatment situations. We assume that $i = 1,\ 2,\dots, n$ are i.i.d. samples. In section 2 and 3, the following linear model is assumed:
\begin{align}
\label{2_1}
y_{i}=t_{i}\beta_{t}+\bld{x}_{i}^{\top}\boldsymbol{\beta}_{x}+u_{i},
\end{align}
where ${\rm E}[U]=0,\ Var(U)=\sigma^2<\infty$ and $\bld{\beta}=\left(\beta_{t},\, \boldsymbol{\beta}^{\top}_{x}\right)^{\top}$. Note that our results can be expanded to more broad situations clearly such as heteroscedasticity of variance, included in interaction terms related to the treatment and covariates, and nonlinear models. However, we consider only simple linear model (\ref{2_1}) to continue the following discussion clearly.

We assume that $U\indep (\bld{Z},\bld{X})$ and $U\not\!\!\!\indep T$. When estimating $\bld{\beta}$ by using OLS, there may be some biases; therefore, the following estimating equation is used (c.f. Hayashi, 2000, Burgess et al., 2017):
\begin{align}
\label{3_1}
\sum_{i=1}^{n}\left(
\begin{array}{c}
h(\bld{z}_{i})\\
\bld{x}_{i}
\end{array}
\right)\left(y_{i}-\left(t_{i}\beta_{t}+\bld{x}_{i}^{\top}\boldsymbol{\beta}_{x}\right)\right)=\bld{0}_{p+m},
\end{align}
where $h(\cdot)$ is $m$-dimensional function ($m\in\{1,2,\dots,K\}$). If the model (\ref{2_1}) is correct, then the estimating equation (\ref{3_1}) becomes
$$
{\rm E}\left[
\left(
\begin{array}{c}
h(\bld{Z})\\
\bld{X}
\end{array}
\right)\left(Y-\left(T\beta_{t}^{0}+\bld{X}^{\top}\boldsymbol{\beta}^{0}\right)\right)
\right]={\rm E}\left[
\left(
\begin{array}{c}
h(\bld{Z})\\
\bld{X}
\end{array}
\right)\right]{\rm E}\left[Y-\left(T\beta_{t}^{0}+\bld{X}^{\top}\boldsymbol{\beta}^{0}\right)\right]=\bld{0}_{p+m},
$$
where the superscript ``$0$" of parameters means the true value of parameters. Therefore, the solution of the estimating equation (\ref{3_1}) becomes the consistent estimator. Under some regularity conditions, the following asymptotic normality holds:
\begin{align}
\label{20200904_1}
\sqrt{n}\left(\hat{\boldsymbol{\beta}}^{(\cdot)}-\boldsymbol{\beta}^{0}\right)\stackrel{L}{\to}N\left(\bld{0}_{p+1},\, \left(\Gamma_{1}\Sigma_{1}^{-1}\Gamma_{1}^{\top}\right)^{-1}\right),
\end{align}
where $H:=h(\bld{Z})$, and
$$
\Sigma_{1}={\rm E}\left[\left(
\begin{array}{c}
HU\\
\bld{X}U
\end{array}
\right)^{\otimes 2}\right]=\sigma^2\left(
\begin{array}{cc}
{\rm E}[H^{\otimes 2}]&{\rm E}[H\bld{X}^{\top}]\\
{\rm E}[\bld{X}H^{\top}]&{\rm E}[\bld{X}\bld{X}^{\top}]
\end{array}
\right),\ \Gamma_{1}=\left(
\begin{array}{cc}
{\rm E}\left[HT\right]&{\rm E}\left[H\bld{X}^{\top}\right]\\
{\rm E}\left[T\bld{X}\right]&{\rm E}\left[\bld{X}\bld{X}^{\top}\right]
\end{array}
\right).
$$
Note that the solution of (\ref{3_1}) are expressed as
$$
\hat{\boldsymbol{\beta}}^{(\cdot)}=\left(\hat{\beta}^{(\cdot)}_{t},\, \left(\hat{\boldsymbol{\beta}}^{(\cdot)}_{x}\right)^{\top}\right)^{\top},
$$
e.g., $\hat{\boldsymbol{\beta}}^{GMM}$ means ordinary GMM estimator when $h(\bld{Z})=\bld{Z}$.

From here, we consider how to select the function $h(\cdot)$ such that the asymptotic variance of (\ref{20200904_1}) is minimized. As describing in Introduction, an allele score, a linear combination of IVs, is well considered in MR (Burgess et al., 2017):
\begin{align}
\label{20200904_2}
h(\bld{z}_{i})=\bld{\gamma}^{\top}\boldsymbol{z}_{i}=\sum_{k=1}^{K}\gamma_{k}z_{ik},
\end{align}
where $\bld{\gamma}\in \mathbb{R}^{K}$. Of course, $h(\bld{z}_{i})$ is also IVs. We denote the solution of (\ref{3_1}) using the equation (\ref{20200904_2}) as $\hat{\boldsymbol{\beta}}^{LC}$ in section 2 (the superscript ``LC'' means ``Linear Combination''). For instance in Burgess et al., 2017, the parameters $\bld{\gamma}$ are estimated as the inverse of standard deviations of each IVs; some ad-hoc procedures are used to estimate $\bld{\gamma}$. From the proposition 1, the solution of a equation $\tilde{\bld{\gamma}}$ can be estimated such that the asymptotic variance of (\ref{20200904_1}) becomes minimum; that is, a IV estimator can be estimated as the ``best'' in the class of (\ref{20200904_2}).
\begin{prop}{$\phantom{}$}\label{te_1}\\
Assume that $\Gamma_{1}>O$. When {\it C.1} and {\it C.2} hold (see Appendix B), the solution of the following equation $\tilde{\bld{\gamma}}$ gives the minimal asymptotic variance of $\hat{\bld{\beta}}^{LC}$:
\begin{align}
\label{3_6}
\left({\rm E}\left[\boldsymbol{Z}^{\otimes 2}\right]-{\rm E}\left[\boldsymbol{Z}\bld{X}^{\top}\right]{\rm E}\left[\bld{X}\bld{X}^{\top}\right]^{-1}{\rm E}\left[\bld{X}\boldsymbol{Z}^{\top}\right]\right)\bld{\gamma}-\left({\rm E}\left[\boldsymbol{Z}T\right]-{\rm E}\left[\boldsymbol{Z}\bld{X}^{\top}\right]{\rm E}\left[\bld{X}\bld{X}^{\top}\right]^{-1}{\rm E}\left[T\bld{X}\right]\right)=\bld{0}_{K}
\end{align}
Then, the asymptotic variance becomes
\begin{align}
\label{20200905_1}
\left(\Gamma_{1}\Sigma_{1}^{-1}\Gamma_{1}^{\top}\right)^{-1}=\sigma^2\left(
\begin{array}{cc}
\Omega_{opt}^{-1}&-\Omega_{opt}^{-1}{\rm E}[T\bld{X}^{\top}]{\rm E}\left[\bld{X}\bld{X}^{\top}\right]^{-1}\\
&{\rm E}\left[\bld{X}\bld{X}^{\top}\right]^{-1}+\sigma^2\Omega_{opt}^{-1}{\rm E}\left[\bld{X}\bld{X}^{\top}\right]^{-1}{\rm E}[T\bld{X}]{\rm E}[T\bld{X}^{\top}]{\rm E}\left[\bld{X}\bld{X}^{\top}\right]^{-1}
\end{array}
\right),
\end{align}
where
\begin{align*}
\Omega_{opt}&=\left({\rm E}\left[\boldsymbol{Z}T\right]-{\rm E}\left[\boldsymbol{Z}\bld{X}^{\top}\right]{\rm E}\left[\bld{X}\bld{X}^{\top}\right]^{-1}{\rm E}\left[T\bld{X}\right]\right)^{\top}\left({\rm E}\left[\boldsymbol{Z}^{\otimes 2}\right]-{\rm E}\left[\boldsymbol{Z}\bld{X}^{\top}\right]{\rm E}\left[\bld{X}\bld{X}^{\top}\right]^{-1}{\rm E}\left[\bld{X}\boldsymbol{Z}^{\top}\right]\right)^{-1}\\
&\hspace{0.5cm}\times\left({\rm E}\left[\boldsymbol{Z}T\right]-{\rm E}\left[\boldsymbol{Z}\bld{X}^{\top}\right]{\rm E}\left[\bld{X}\bld{X}^{\top}\right]^{-1}{\rm E}\left[T\bld{X}\right]\right)
\end{align*}
\end{prop}

The proof of proposition \ref{te_1} is shown in Appendix D. From the result of proposition \ref{te_1}, we can estimate the weights of an allele score so that the asymptotic variance of $\hat{\bld{\beta}}^{LC}$ becomes minimal. By the way, as a well-known fact, the GMM has the semiparametric efficiency; the asymptotic variance of (\ref{20200904_1}) is minimized when the function $h(\cdot)$ is selected as $h(\bld{Z})=\bld{Z}$. From the next theorem, we can derive the important conclusion that $\hat{\bld{\beta}}^{LC}$ also has the semiparametric efficiency. In other word, $\hat{\bld{\beta}}^{LC}$ becomes the ``best'' in the class of the solution of (\ref{3_1}).
\begin{theo}{$\phantom{}$}\label{te_2}\\
When {\it C.1} and {\it C.2} hold, the asymptotic variance of $\hat{\bld{\beta}}^{GMM}$ is the same as (\ref{20200905_1}).
\end{theo}\noindent
The proof of theorem \ref{te_2} is shown in Appendix D. The solution of (\ref{3_6}) $\tilde{\bld{\gamma}}$ gives the minimal asymptotic variance of $\hat{\boldsymbol{\beta}}^{(\cdot)}$; it is one of a different point from GMM. By using the feature, we can propose the valid IVs selection method in the next section.

\section{Situations where some instrumental variables are invalid}
In the previous section, we discussed the situation where all $\bld{Z}$ are valid IVs, but in the case of a large number of IVs such as genetic variants, there are risks that estimating biased causal effects by including some invalid IVs. In this section, we discuss the situation where some of candidates of IVs $(Z_{\ell+1},\dots,Z_{K})$ do not satisfy at least IV condition 2) or 3) explained at Introduction; in other words, $(Z_{\ell+1},\dots,Z_{K})$ are invalid IVs.

Invalid IVs do not satisfy IV condition 2) or 3); mathematically denoted as $Z_{k}\notindep U,\ k\in\{\ell+1,\dots,K\}$ in this paper. For instance, $U$ can be written as $U=g(Z_{k})+\zeta$, where ${\rm E}[g(Z_{k})]={\rm E}[\zeta]=0$, $Var(g(Z_{k}))<\infty$, $Var(\zeta)<\infty$, and $\zeta\indep\left(\bld{Z},\bld{X}\right)$. Under this folmulation, (\ref{2_1}) becomes
$$
y_{i}=t_{i}\beta_{t}+\bld{x}_{i}^{\top}\boldsymbol{\beta}_{x}+g(z_{ik})+\zeta.
$$
Under the violation of IV condition 2), $U$ is considered as an ``unmeasured phenotype" affected by horizontal pleiotropy. Under the violation of IV condition 3), $U$ is considered as an ``unmeasured covariate" derived from unaware population structure. Since $U$ is an unobserved variable, the nature of the variable itself cannot be directly examined from the data; in this paper we consider to use Negative Control Outcomes (NCOs) as auxiliary variables.

At first, we introduce mathematically definition of NCOs (see Appendix A also).

\begin{defi}{$\phantom{}$}\\
\label{def1}
Negative Control Outcome (NCO) $M\in\mathbb{R}$ are assumed to be the variable sasisfying the following conditions:
\begin{enumerate}
\item ${\rm E}\left[M\right]=0$
\item $M=M(U)+\varepsilon$,\ \ $\varepsilon\indep(T,\bld{X},\bld{Z}),\, {\rm E}\left[\varepsilon\right]=0,\, Var(\varepsilon)<\infty$
\item $|{\rm E}\left[Z_{k}M\right]|>w>0$, if $k\in\{\ell+1,\dots,K\}$
\end{enumerate}
\end{defi}\noindent
{\it 1}. is not an essential assumption, but to simplify the discussion below. {\it 2}. is an assumption for detect valid IVs from the candidates. Under {\it 1} and {\it 2}., when $k\in\{1,\dots,\ell\}$,
$$
{\rm E}\left[Z_{k}M\right]={\rm E}\left[M(U){\rm E}\left[Z_{k}|U\right]\right]={\rm E}\left[M(U)\right]{\rm E}\left[Z_{k}\right]=0.
$$
{\it 3}. expresses the relationship between invalid IVs and NCO. Since invalid IVs are related to the unobserved variable, the assumption is appropriate. By utilizing from {\it 1} to {\it 3}., it is possible to identify valid and invalid IVs by NCO. For {\it 2}., the similar assumptions are made in Miao and Tchetgen Tchetgen, 2018 (auxiliary variables and IVs are conditionally independent given unobserved covariates ($M\indep \bld{Z}|U$)), and NCO is, for this purposes, only related to unobserved variables; {\it 2}. is considered to be natural. {\it 3}. is a necessary assumption since assuming the general formulation of $M(U)$ in {\it 2}. assumes a general system of functions. If a linear model can be assumed between $M$ and $U$, then the assumption of covariance between $M$ and $U$ turns out to be sufficient. Note that NCO has been usually used in epidemiological studies as a way to check for the effects of unmeasured covariates (for instance, see Miao and Tchetgen Tchetgen, 2018, p.7). This is achieved from the feature not related to IVs and treatments directly; only through the unmeasured covariates. Therefore, the assumption is slightly different from definition \ref{def1}; definition \ref{def1} is broader than ordinary NCOs.

\subsection{The proposed method and asymptotic properties}
From here, we update the proposed method introduced in the previous section, and we propose the new method different from existing methods to estimate causal effects while selecting valid IVs. Note that to simplify the following discussions, we assume ${\rm E}\left[\bld{X}\bld{X}^{\top}\right]={\rm I}_{p}$, but the assumption is not essential. In the proposed method, the weights $\bld{\gamma}$ can be estimated as the solution of the equation (\ref{3_6}) with changing to sample means. When there exists both valid and invalid IVs in the candidates, the weights related to valid IVs $\bld{\gamma}_{val}=\left(\gamma_{0},\dots,\gamma_{\ell}\right)^{\top}$ would like to be estimated from the estimation equation; the other weights $\bld{\gamma}_{inv}=\left(\gamma_{\ell+1},\dots,\gamma_{K}\right)^{\top}$ would like to be estimated as $0$, or convergence sequences to $0$. As the estimator of $\bld{\gamma}$, the following estimating equation can be considered:
\begin{align}
\label{3_9}
\sum_{i=1}^{n}\underbrace{\left(
\begin{array}{ccc}
\left(Z_{i1}^2-\sum_{j=1}^{p}\overline{Z_{1}X_{j}}^2\right)I_{\tau 1}+\kappa_{1}\bar{I}_{\tau 1}&\cdots&\left(Z_{i1}Z_{iK}-\sum_{j=1}^{p}\overline{Z_{1}X_{j}}\, \overline{Z_{K}X_{j}}\right)I_{\tau 1}I_{\tau K}\\
&\ddots&\vdots\\
&&\left(Z_{iK}^2-\sum_{j=1}^{p}\overline{Z_{K}X_{j}}^2\right)I_{\tau K}+\kappa_{1}\bar{I}_{\tau K}
\end{array}
\right)}_{=:A_{i}}\bld{\gamma}&\nonumber\\
&\hspace{-12.5cm}=\sum_{i=1}^{n}\underbrace{\left(
\begin{array}{c}
\left(Z_{i1}T_{i}-\sum_{j=1}^{p}\overline{Z_{1}X_{j}}\, \overline{TX_{j}}\right)I_{\tau 1}+\kappa_{2n}\bar{I}_{\tau 1}\\
\vdots\\
\left(Z_{iK}T_{i}-\sum_{j=1}^{p}\overline{Z_{K}X_{j}}\, \overline{TX_{j}}\right)I_{\tau K}+\kappa_{2n}\bar{I}_{\tau K}
\end{array}
\right)}_{=:\bld{b}_{i}},
\end{align}
where $\bld{\gamma}=\left(\bld{\gamma}_{val}^{\top},\bld{\gamma}_{inv}^{\top}\right)^{\top},\ \kappa_{1}\in \mathbb{R}, \kappa_{2n}=o\left(\frac{1}{\sqrt{n}}\right)$,
$$
\overline{Z_{k}X_{j}}=\frac{1}{n}\sum_{i=1}^{n}Z_{ik}X_{ij},\ \overline{TX_{j}}=\frac{1}{n}\sum_{i=1}^{n}T_{i}X_{ij}
$$
\begin{align}
\label{3_11}
I_{\tau k}=\Phi_{\tau}\left(\hat{w}_{k}+w\right)\Phi_{\tau}\left(w-\hat{w}_{k}\right),\ \bar{I}_{\tau k}=1-I_{\tau k},
\end{align}
\begin{align}
\label{3_12}
I^{0}_{\tau k}=\Phi_{\tau}\left(w^{0}_{k}+w\right)\Phi_{\tau}\left(w-w^{0}_{k}\right),\ \bar{I}^{0}_{\tau k}=1-I^{0}_{\tau k},
\end{align}
\begin{align}
\label{3_13}
\hat{w}_{k}=\frac{1}{n}\sum_{i=1}^{n}Z_{k i}M_{i},\ w_{k}^{0}={\rm E}\left[Z_{k}M\right],
\end{align}
and $\Phi_{\tau}(\cdot)$ is CDF of $N(0,\tau^2)$. (\ref{3_11}), (\ref{3_12}) are estimators and the true values of the {\it smooth weight function} (c.f. Yang and Ding, 2017, Fig.1), respectively. From here, we confirm formulas (\ref{3_9})-(\ref{3_13}). At first, we assume $Z_{k}$ is valid.  Then, ${\rm E}\left[Z_{k}M\right]= 0$, and
$$
I^{0}_{\tau k}=\Phi_{\tau}\left(0+w\right)\Phi_{\tau}\left(w-0\right)\to 1\ \ \ (\tau\searrow 0).
$$
Therefore, it is expected that
$$
I_{\tau k}=\Phi_{\tau}\left(\hat{w}_{k}+w\right)\Phi_{\tau}\left(w-\hat{w}_{k}\right)\stackrel{P}{\to} 1\ \ \ (n\to \infty,\, \tau\searrow 0).
$$
Whereas, we assume $Z_{k'}$ is invalid. Then, $\left|{\rm E}\left[Z_{k'}M\right]\right|> w$, and
$$
I^{0}_{\tau k'}=\Phi_{\tau}\left(w_{k'}^{0}+w\right)\Phi_{\tau}\left(w-w_{k'}^{0}\right)\to 0\ \ \ (\tau\searrow 0).
$$
Therefore, it is expected that
$$
I_{\tau k'}=\Phi_{\tau}\left(\hat{w}_{k'}+w\right)\Phi_{\tau}\left(w-\hat{w}_{k'}\right)\stackrel{P}{\to} 0\ \ \ (n\to \infty,\, \tau\searrow 0).
$$
In summary, about (\ref{3_9}), it is expected that weights $\bld{\gamma}_{val}$ of valid IVs and $\bld{\gamma}_{inv}$ of invalid IVs are consistent asymptotically with solutions of following estimating equations respectively:
\begin{align}
\label{3_18}
\sum_{i=1}^{n}\underbrace{\left(
\begin{array}{ccc}
Z_{1i}^2-\sum_{j=1}^{p}\overline{Z_{1}X_{j}}^2&\cdots&Z_{1i}Z_{\ell i}-\sum_{j=1}^{p}\overline{Z_{1}X_{j}}\, \overline{Z_{\ell}X_{j}}\\
&\ddots&\vdots\\
&&Z^2_{\ell i}-\sum_{j=1}^{p}\overline{Z_{\ell}X_{j}}^2
\end{array}
\right)}_{=:\check{A}_{i}}\bld{\gamma}_{val}&\nonumber\\
&\hspace{-8.5cm}=\sum_{i=1}^{n}\underbrace{\left(
\begin{array}{c}
\left(Z_{i1}T_{i}-\sum_{j=1}^{p}\overline{Z_{1}X_{j}}\, \overline{TX_{j}}\right)\\
\vdots\\
\left(Z_{i\ell}T_{i}-\sum_{j=1}^{p}\overline{Z_{\ell}X_{j}}\, \overline{TX_{j}}\right)
\end{array}
\right)}_{=:\check{\bld{b}}_{i}},
\end{align}
$$
\kappa_{1}\times diag(1,\dots,1)\bld{\gamma}_{inv}=0
$$

To prove the above expectations, we confirm properties of $\bld{\gamma}$ through the following two steps:
\begin{description}
\item [Step 1)] Derive an asymptotic equivalent random variable to (\ref{3_9})
\item [Step 2)] By using random variable derived at Step 1), confirming mathematical properties of the following formula:
$$
\hat{\bld{\gamma}}=\left(\sum_{i=1}^{n}A_{i}\right)^{-1}\sum_{i=1}^{n}\bld{b}_{i}
$$
\end{description}
Step 1) and 2) is acheived by Lemma 1 and Proposition 2 described in appendix C. By using $\hat{\bld{\gamma}}$, an estimating equation for $\bld{\beta}$ can be constructed:
\begin{align}
\label{3_19}
\sum_{i=1}^{n}\left(
\begin{array}{c}
\hat{\bld{\gamma}}^{\top}\bld{z}_{i}\\
\bld{x}_{i}
\end{array}
\right)\left(y_{i}-\left(t_{i}\beta_{t}+\bld{x}_{i}^{\top}\boldsymbol{\beta}_{x}\right)\right)=\bld{0}_{p+1}.
\end{align}
Regarding $\bld{\beta}$, the following propertiy holds:
\begin{theo}{$\phantom{}$}\\
\label{te3_2}
When {\it C.1} and {\it C.2}, some regularity conditions described in the appendix hold, and $\hat{\bld{\gamma}}$ is the solution of (\ref{3_9}). Then, $\hat{\bld{\beta}}^{LC}$ the solution of (\ref{3_19}) has the following asymptotical property:
\begin{align}
\label{3_20}
\sqrt{n}\left(\hat{\bld{\beta}}^{LC}-\bld{\beta}^{0}\right)\stackrel{L}{\to}N\left(\bld{0}_{p+1},\, \left(\Gamma_{2}\Sigma_{2}^{-1}\Gamma_{2}^{\top}\right)^{-1}\right),
\end{align}
where $\bld{\gamma}^{0}=\left((\bld{\gamma}^{0}_{val})^{\top},(\bld{\gamma}_{inv}^{0})^{\top}\right)^{\top}\equiv \left((\bld{\gamma}^{0}_{val})^{\top},\bld{0}_{K-\ell}^{\top}\right)^{\top}$ and
$$
\Sigma_{2}=\sigma^2{\rm E}\left[\left(
\begin{array}{c}
\left(\bld{\gamma}^{0}\right)^{\top}\bld{Z}\\
\bld{X}
\end{array}
\right)^{\otimes 2}\right],\ \Gamma_{2}=\left(
\begin{array}{cc}
{\rm E}\left[\left(\bld{\gamma}^{0}\right)^{\top}\bld{Z}T\right]&{\rm E}\left[\left(\bld{\gamma}^{0}\right)^{\top}\bld{Z}\bld{X}^{\top}\right]\\
{\rm E}\left[T\bld{X}\right]&{\rm I}_{p}
\end{array}
\right).
$$
\end{theo}\noindent
The proof of {theorem \ref{te3_2} is shown in Appendix D. Theorem \ref{te3_2} shows that $\hat{\bld{\beta}}^{LC}$ can be satisfied semiparametric efficiency when there are some invalid IVs; therefore the conclusions derived in the previous section hold. One of the important points in theorem \ref{te3_2} is that the IV estimator does not affect the asymptotic variance when applying the proposed method. This is because the variability of valid IVs is independent of unobserved covariates, and the variability of invalid IVs becomes $0$ by selecting $\kappa_{2n}=o(1/\sqrt{n})$.

In this paper, we propose the following procedures to estimate consistent causal effects when NCO can be obtained:
\begin{enumerate}
\item $\hat{\bld{\gamma}}$ is estimated as the solution of (\ref{3_9}).
\item By using $\hat{\bld{\gamma}}$ in (\ref{3_19}), we obtain an consistent estimator of causal effects.
\end{enumerate}
As obviously, to implement our proposed method, we have to decide tuning parameters $(\kappa_{1},\kappa_{2n},\tau,w)$. The handling of these parameters are described in appendix F. The above procedures are implemented to the devided two samples; our proposed methods can be applied in both one-sample and two-sample mendelian randomization. Also, when there are no NCO but at least one of the candidates of valid IVs can be detected, the valid IVs can be used as auxiliary variables; the same situation as Liao, 2013 and DiTraglia, 2016 (see Appendix E).

\subsection{Binary outcomes extention}
By the previous subsection, we assume continuous outcomes (more precisely, linear relationship between an outcome and a treatment); however, it is common to apply nonlinear models such as binary outcomes in biometrics (i.e., $\mathcal{Y}=\{0,1\}$). In this subsection, we would like to consider the expansion of the above considered model.

Applying to the binary outcomes, log linear model or logistic regression model are commonly used: 
\begin{align}
{\rm E}\left[Y|t_{i},\bld{x}_{i},u_{i}\right]&=\exp\left\{t_{i}\beta_{t}+\bld{x}_{i}^{\top}\boldsymbol{\beta}_{x}+u_{i}\right\},\label{llin}\\
{\rm E}\left[Y|t_{i},\bld{x}_{i},u_{i}\right]&={\rm expit}\left\{t_{i}\beta_{t}+\bld{x}_{i}^{\top}\boldsymbol{\beta}_{x}+u_{i}\right\}.\label{logis}
\end{align}
In the following discussion, (\ref{llin}) is only considered. To estimate the parameters, the following estimating equation is considered:
$$
\sum_{i=1}^{n}\left(
\begin{array}{c}
h(\bld{z}_{i})\\
\bld{x}_{i}
\end{array}
\right)\left(y_{i}-{\rm exp}\left\{t_{i}\beta_{t}+\bld{x}_{i}^{\top}\boldsymbol{\beta}_{x}\right\}\right)=\bld{0}_{p+1}.
$$
However, the estimator may have some biases. This is because
$$
{\rm E}\left[\left(
\begin{array}{c}
h(\bld{Z})\\
\bld{X}
\end{array}
\right)\left(Y-{\rm exp}\left\{T\beta_{t}^{0}+\bld{X}^{\top}\boldsymbol{\beta}^{0}_{x}\right\}\right)\right]\neq\bld{0}_{p+1}.
$$
To solve the problem, we consider that (\ref{llin}) can be approximated by the following model:
\begin{align}
\label{3_22}
{\rm E}\left[Y|t_{i},\bld{x}_{i},u_{i}\right]\approx{\rm exp}\left\{t_{i}\beta_{t}+\bld{x}_{i}^{\top}\boldsymbol{\beta}_{x}\right\}+u_{i},
\end{align}
Obviously, the estimating equation based on (\ref{3_22}) derive unbiased estimator. The model is ``approximation of binary outcome models" (Johnston et al., 2008 and Clarke and Windmeijer, 2012). As mentioned in Clarke and Windmeijer (2012), the approximation (\ref{3_22}) is somewhat reasonable in the sense of a first-order approximation, or when a variance of unobserved variables is only small. In this sense, the approximation such as (\ref{3_22}) is acceptable to apply our proposed method, and the model is used for the following real data analysis.

\section{Simulations}
In this section, we confirm properties of our proposed method and compare with the method put forward by Liao, 2013, DiTraglia, 2016, and some methods used in a mendelian randomization. Our simulation is constructed by two situations:
\begin{enumerate}
\item Confirming the performance of our proposed method and the previous methods through under the setting of Liao, 2013.
\item Confirming the performance of our proposed method and the previous methods through under the setting of Gkatzionis et al. (2021); more realistic setting for a mendelian randomization.
\end{enumerate}
Through the situation 1., we show that 1) our proposed method has the same asymptotic variance as GMM, 2) pre-specified valid IVs is not necessary to our proposed method when there are some NCOs. Through the situation 2., we show that our proposed method works well under some mendelian randomization settings when there are some NCOs. To confirm these properties, we summarize descriptive statistics of estimates for each procedure. The number of iterations for simulations is 1,000. The details of simulation under the setting of Liao, 2013 is appeared in appendix G.

In simulation 1, it is appeared that our proposed method has well efficiency, and need not to detect some valid IVs before analyses. In this simulation, we consider more realistic settings. Specifically, we consider the situation where there are some genetic variants; however, we cannot detect which genetic variants are valid IVs. Therefore, there is a possibility that we estimate a biased causal effect. The simulation setting is as follows (refer to Gkatzionis et al., 2021):
\begin{description}
\item{{\bf Candidates of IVs}}
$$
Z_{ik}\stackrel{i.i.d.}{\sim}Binom(2,p_{zk}),\ \ p_{zk}\sim Unif(0.1,0.9),\ \ k=1,2,\dots,100
$$
\item{{\bf Unmeasured covariates}}
\begin{align*}
U_{i}&=(Z_{i1}-2p_{z1},\dots,Z_{i100}-2p_{z100})\bld{\alpha}+\varepsilon_{ui},\ \varepsilon_{ui}\sim N(0,0.1^2),\\
\bld{\alpha}^{\top}&=(\underbrace{0,\dots,0}_{valid\ IVs},\underbrace{\tilde{\bld{\alpha}}^{\top}}_{invalid\ IVs}),\ \tilde{\alpha}_{k}\sim N(0.4,0.1^2)
\end{align*}
\end{description}
Note that we consider two settings. At first, valid IVs are majority situation: valid IVs are 70 and invalid IVs are 30. Secondary, invalid IVs are majority situation: valid IVs are 30 and invalid IVs are 70. Also, $\frac{1}{n}\sum_{i=1}^{n}U_{i}\stackrel{P}{\to}0$.
\begin{description}
\item{{\bf Treatment}}
$$
T_{i}=(Z_{i1},\dots,Z_{i100})\bld{\beta}+U_{i},\ \ {\beta}_{k}=0.5+|\tilde{\beta}_{k}|,\ \ \tilde{\beta}_{k}\sim N(0,0.5^2)
$$
\item{{\bf Outcome}}
$$
Y_{ti}=1+0.3T_{i}+U_{i}
$$
\item{{\bf Negative Control Outcome}}
$$
M_{i}=1+\alpha_{mu}U_{i}+\varepsilon_{mi},\ \ \varepsilon_{mi}\sim N(0,1)
$$
\begin{enumerate}
\item Valid IVs are majority situation: $\alpha_{mu}=0.8$
\item Invalid IVs are majority situation: $\alpha_{mu}=0.5$\\
$\Rightarrow$Correlation between a NCO and an unmeasured covariate becomes approximately 0.75 (i.e. we consider a strong NCO situation).
\end{enumerate}
\end{description}
In this simulation, the proposed estimator is compared with the ordinary GMM, the well-known robust methods (Inverse-variance weighting (IVM) and MR-Egger), and the novel robust methods (a median based estimator (Bowden et al., 2016) and a mode based estimator (Hartwig et al., 2017)). Note that only single sample Mendelian randomization is considered in this simulation.

The simulation results are summarized in figures. The estimated causal effects are summarized in Figure 3 (included in tables also), and Figure 6 and 7 in appendix G. Our proposed method has good performance compared with the other four methods. The mode based estimator works well when valid IVs are majority situation; the median based estimator somewhat works also. Whereas, the well-known robust methods do not work when both situations. Regarding invalid IVs are majority situation, this is the situation where Zero Modal Pleiotropy Assumption (see Hartwig et al., 2017) is not valid, there is an obvious bias except for our proposed method. Whereas, our proposed method also has a bias, but magnitude is relatively small. Therefore, our proposed method works well when there are only the small number of valid IVs, and there is a suitable NCO. Focus once again on the proposed method. The method has somewhat bias. This is derived from the ``careless selection" of invalid IVs. In Figure 6 and 7, this is one of the example of selection of IVs, some invalid IVs are selected incorrectly. As mentioned in appendix F, the power (``all covariances between invalid IVs and the NCO exceed the cut-off point $w$") may be insufficient.

\section{Conclusions and Future Works}
In this paper, we proposed the novel IV estimator which use a linear combination of IVs. When there are some invalid IVs, our proposed method can select the valid IVs by using a negative control outcome or some auxiliary variable. Whether selecting valid IVs or not, we showed that our proposed estimator has the same efficiency of the generalized methods of moments estimator. We confirm performances of our proposed method and some previous methods through simulations. Our proposed method also works; especially it is remarkable point that the method can be applied where there is only the small number of valid IVs.

We believe that our proposed estimator has important impact on the biometrics and related fields, specifically in Mendelian Randomization. As we mentioned in Introduction, there are many works related to IV methods, however, there may be some applicational problems since a couple of important assumptions are necessary for some previous methods (the number of valid IVs, the feature of IV candidates, model assumptions, ...). Our results show that by using an auxiliary variable such as a negative control outcome, we can relax such assumptions; the results may become one of the key conclusion to solve the important IV problems globally. On the other hand, some auxiliary variables are necessary to our proposed method (in other words, ``using some auxiliary variable" is an important assumption), but we suppose this is correct intuitively. Since we cannot observe unmeasured variables, we have to observe a fluctuation of some proxy variables; this is the role of an auxiliary variable. The ``proxy variable" based estimator is recently considered well (Miao and Tchetgen Tchetgen, 2018 and Cui et al., 2020).

Our proposed method has many interesting points, but there are also many future works. First of all, our discussions assume the correct outcome model. If we misspecify the model, of course our proposed estimator does not have even the consistency. Okui et al., 2012 and Ogburn et al, 2015 have proposed a doubly robust estimator in the sense that we only need to specify the correct model either an outcome or instrumental variables. Our proposed method may be extended as having double robustness by applying their ideas. Secondly, our proposed method assume continuous or binary outcomes. In biometrics and related fields, time-to-event type outcomes are well considered. Recently, some methods overcoming the problem of unobserved covariate have been proposed in recent years (e.g., Tchetgen Tchetgen et al., 2015, Kjaersgaard and Parner, 2016, Mart\'{i}nez-Camblor et al., 2019, and Orihara, 2022). Since our proposed method can be extended to apply various type of data, the IV problems may be overcome in the situation. Thirdly, our proposed method needs to specify an auxiliary variable related to unobserved variables. In other words, the unobserved variables related to the used auxiliary variable is only detected. To solve the problem, using multiple auxiliary variables as many as possible become one of the solutions. We have to formulate the situation, and confirm the properties to understand our proposed method more clearly. Finally, in our paper, we assume the fixed number of instrumental variables. In the econometrics, the situation where the number of instrumental variables also increase when sample size increase is well considered (e.g. Bekker, 1994 and Okui, 2011). It is well known that more correct confidence intervals are derived by taking into consideration of the number of IVs. For more accurate inference, we need to update our proposed method.

\begin{singlespace}

\end{singlespace}

\appendix

\begin{landscape}
\begin{figure}[h]
\begin{center}
\begin{tabular}{c}
\includegraphics[width=24cm]{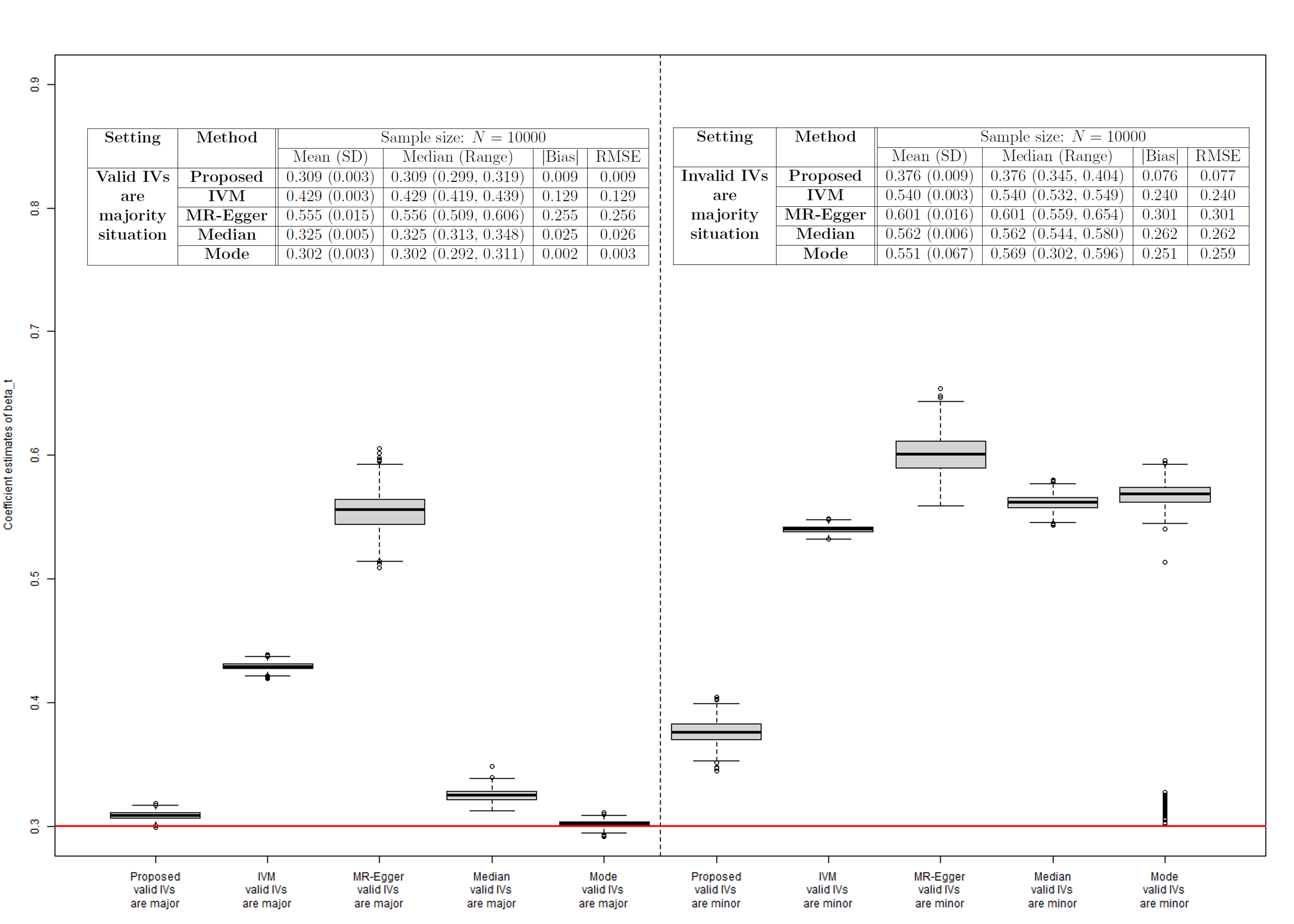}
\end{tabular}
\label{fig3}
\caption{Summary of estimators for causal effects}
\end{center}
\end{figure}
\end{landscape}

\section{Supplementary figures}
\begin{figure}[h]
\begin{center}
\begin{tabular}{c}
\includegraphics[width=8.5cm]{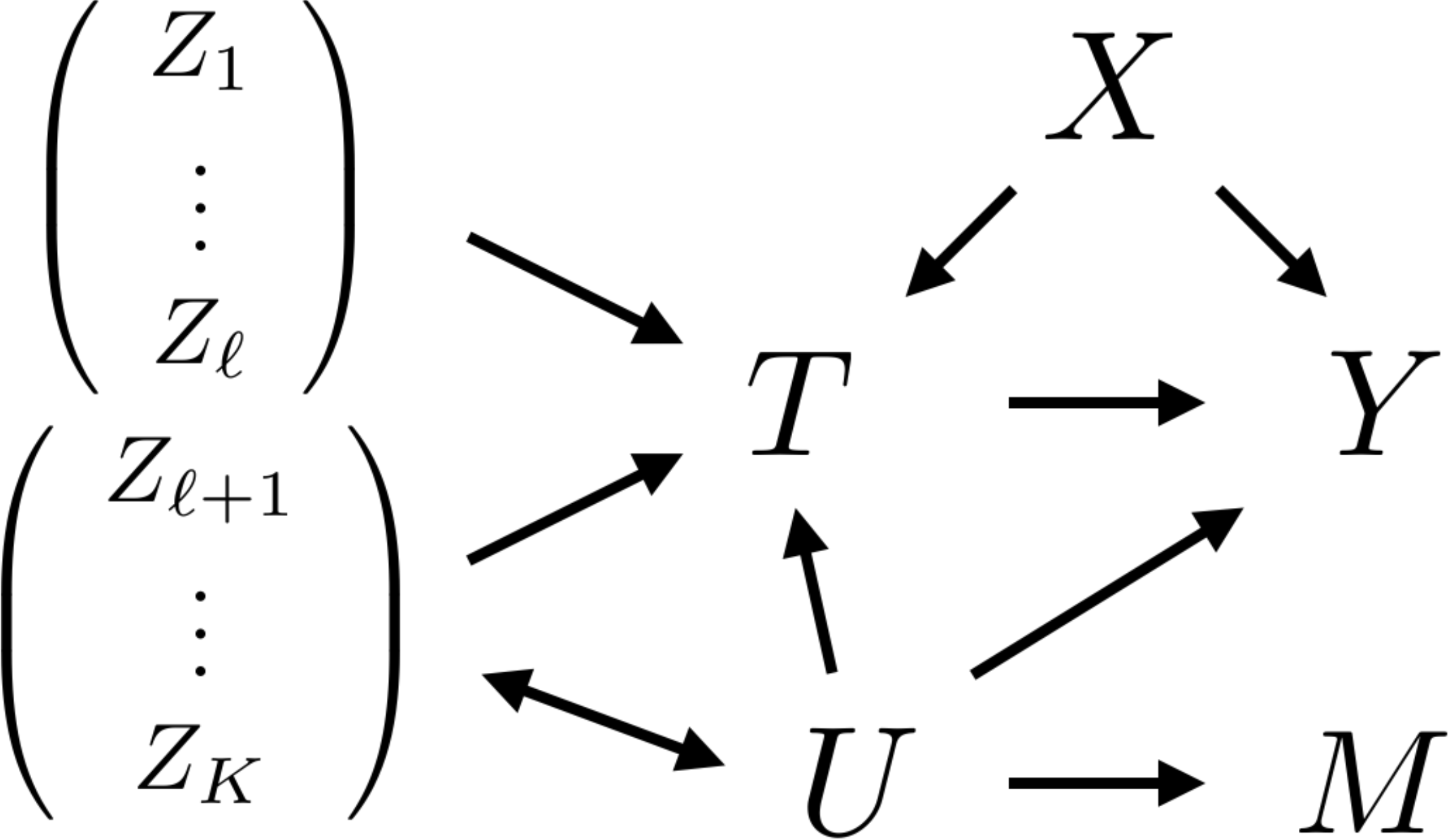}
\end{tabular}
\caption{Relationship of variables with NCO}
\end{center}
\end{figure}

\newpage
\section{Important regularity conditions}
Throughout our papers, the following two regularity conditions are important:
\begin{description}
\item{{\large C.1}}
$$
{\rm E}\left[\boldsymbol{Z}^{\otimes 2}\right]-{\rm E}\left[\boldsymbol{Z}\bld{X}^{\top}\right]{\rm E}\left[\bld{X}\bld{X}^{\top}\right]^{-1}{\rm E}\left[\bld{X}\boldsymbol{Z}^{\top}\right] >O
$$
\item{{\large C.2}}
$$
{\rm E}\left[\boldsymbol{Z}T\right]-{\rm E}\left[\boldsymbol{Z}\bld{X}^{\top}\right]{\rm E}\left[\bld{X}\bld{X}^{\top}\right]^{-1}{\rm E}\left[T\bld{X}\right]\neq\bld{0}
$$
\end{description}
{\it C.1} means valid relationship between $\bld{Z}$ and $\bld{X}$; it shows only relationship of variables. Whereas, {\it C.2} is regarded as a kind of IV condition 1). If
$$
{\rm E}\left[\boldsymbol{Z}T\right]-{\rm E}\left[\boldsymbol{Z}\bld{X}^{\top}\right]{\rm E}\left[\bld{X}\bld{X}^{\top}\right]^{-1}{\rm E}\left[T\bld{X}\right]=\bld{0},
$$
i.e., $\bld{\gamma}^{0}=\bld{0}$, a linear combination of IVs does not work. Therefore, an IV estimator cannot be constructed.

\section{Important Lemma and Propositions}

When assuming $\tau\equiv\tau_{n}\searrow 0$, the following lemma can be obtained.
\begin{lemm}{$\phantom{}$}\\
\label{ho3_2}
Considering the following equation:
\begin{align}
\label{20200914_1}
\frac{1}{n}\sum_{i=1}^{n}\left(
\bld{b}_{i}-A_{i}\bld{\gamma}
\right)-\frac{1}{n}\sum_{i=1}^{n}\left\{
\left(
\begin{array}{c}
\check{\bld{b}}_{i}\\
\kappa_{2n}\bld{1}_{K-\ell}
\end{array}
\right)-\left(
\begin{array}{cc}
\check{A}_{i}&O_{\ell\times K-\ell}\\
O_{\ell\times K-\ell}^{\top}&\kappa_{1}{\rm I}_{K-\ell}
\end{array}
\right)\bld{\gamma}
\right\}&\nonumber\\
&\hspace{-11.5cm}=\frac{1}{n}\sum_{i=1}^{n}\underbrace{\left(
\bld{b}_{i}-\left(
\begin{array}{c}
\check{\bld{b}}_{i}\\
\kappa_{2n}\bld{1}_{K-\ell}
\end{array}
\right)\right)}_{=:\bld{c}_{i1}}
-\frac{1}{n}\sum_{i=1}^{n}\underbrace{\left(
A_{i}-\left(
\begin{array}{cc}
\check{A}_{i}&O_{\ell\times K-\ell}\\
O_{\ell\times K-\ell}^{\top}&\kappa_{1}{\rm I}_{K-\ell}
\end{array}
\right)
\right)\bld{\gamma}}_{=:\bld{c}_{i2}(\bld{\gamma})}.
\end{align}
A sufficient condition that the following equations:
\begin{align}
\label{3_14}
\frac{1}{n}\sum_{i=1}^{n}c_{i1k}=o_{p}\left(\frac{1}{\sqrt{n}}\right),\ \frac{1}{n}\sum_{i=1}^{n}c_{i2k}(\bld{\gamma})=o_{p}\left(\frac{1}{\sqrt{n}}\right)
\end{align}
are satisfied for $\forall\bld{\gamma}$ and each $k$ is
\begin{align}
\label{3_15}
\left(\frac{\left|w-\hat{w}_{k}\right|}{\tau_{n}}\right)^{-1}\exp\left\{-\frac{1}{2}\frac{\left(w-\hat{w}_{k}\right)^2}{\tau_{n}^2}\right\}=o_{p}\left(\frac{1}{\sqrt{n}}\right).
\end{align}
\end{lemm}\noindent
{\bf Lemma \ref{ho3_2}.} gives the sufficient order to satisfy (\ref{3_14}), but (\ref{3_15}) is hard to interpret. We assume that
$$
b_{n}=\sqrt{n}\left(\frac{\left|w-\hat{w}_{k}\right|}{\tau_{n}}\right)^{-1}\exp\left\{-\frac{1}{2}\frac{\left(w-\hat{w}_{k}\right)^2}{\tau_{n}^2}\right\}.
$$
\begin{itemize}
\item When $(\left|w-\hat{w}_{k}\right|)/\tau_{n}=O_{p}(n)$,
$$
b_{n}=O_{p}\left(\frac{1}{\sqrt{n}}\right)\times O_{p}\left(\exp\left(-n^2\right)\right)=o_{p}(1)
$$
\item When $(\left|w-\hat{w}_{k}\right|)/\tau_{n}=O_{p}(\sqrt{\log{n}})$,
$$
b_{n}=O_{p}\left(\frac{\sqrt{n}}{\sqrt{\log{n}}}\right)\times O_{p}\left(n^{-\frac{1}{2}}\right)=o_{p}(1)
$$
\item When $(\left|w-\hat{w}_{k}\right|)/\tau_{n}=O_{p}(\sqrt{\log{\log{n}}})$,
$$
b_{n}=O_{p}\left(\frac{\sqrt{n}}{\sqrt{\log{\log{n}}}}\right)\times O_{p}\left(\left(\log n\right)^{-\frac{1}{2}}\right)=O_{p}\left(\frac{\sqrt{n}}{\sqrt{\log{n\log{n}}}}\right)
$$
\end{itemize}
From the above, one of a limit of an order satisfied (\ref{3_15}) is $(\left|w-\hat{w}_{k}\right|)/\tau_{n}=O_{p}(\sqrt{\log{n}})$. Then, since $\hat{w}_{k}$ is a sample mean, a lower limit of an order for $\tau_{n}$ is
$$
\tau_{n}^2=O_{p}\left(\frac{1}{\log n}\right).
$$

From {\bf Lemma \ref{ho3_2}.}, the {\it root-n consistency} of $\hat{\bld{\gamma}}$ is satisfied.
\begin{prop}{$\phantom{}$}\\
\label{te3_1}
When {\it C.1} holds, under the conditions of {\bf Lemma \ref{ho3_2}.},
\begin{align}
\hat{\bld{\gamma}}=\left(\sum_{i=1}^{n}A_{i}\right)^{-1}\sum_{i=1}^{n}\bld{b}_{i}\stackrel{P}{\to}\left(
\begin{array}{c}
{\rm E}\left[\check{A}\right]^{-1}{\rm E}\left[\check{\bld{b}}\right]\\
\bld{0}_{K-\ell}
\end{array}
\right)=\left(
\begin{array}{c}
\bld{\gamma}_{val}^{0}\\
\bld{\gamma}_{inv}^{0}
\end{array}
\right).
\end{align}
Also, if $\left|{\rm E}\left[\left(\check{\bld{b}}-\check{A}\bld{\gamma}_{val}^{0}\right)^{\otimes 2}\right]\right|<\infty$, then each $k$,
$$
\hat{\gamma}_{val,\, k}-\gamma^{0}_{val,\, k}=O_{p}\left(\frac{1}{\sqrt{n}}\right),\ \hat{\gamma}_{inv,\, k}=o_{p}\left(\frac{1}{\sqrt{n}}\right).
$$
\end{prop}\noindent
Note that $\bld{\gamma}_{val}^{0}$ is the solution of (2.5), and $\bld{\gamma}_{inv}^{0}=\bld{0}_{K-\ell}$. In other words, the weights related to invalid IVs are not used when estimating $\hat{\bld{\beta}}^{LC}$. 

\section{Proofs}
\subsection{Proof of Proposition 1}\noindent
We use the following lemma that is well-known conclusion (c.f. Harville, 2006) when showing the {\bf Proposition 1}.
\begin{lemm}{$\phantom{}$}\\
\label{ho3_1}
$A$ means $m\times m$ non-singular matrix, $\boldsymbol{d}$ means $m$-dimension vector. $\forall \boldsymbol{x}\, \backslash\, \{0\}\in \mathbb{R}^{m}$,
$$
\frac{\boldsymbol{x}^{\top}A\boldsymbol{x}}{\left(\boldsymbol{d}^{\top}\boldsymbol{x}\right)^2}\geq \left(\boldsymbol{d}^{\top}A^{-1}\boldsymbol{d}\right)^{-1},
$$
the equation holds when $\boldsymbol{x}\propto A^{-1}\boldsymbol{d}$.
\end{lemm}
From here, we prove {\bf Proposition 1}. At first, we calculate $\Gamma_{1}^{-1}$: 
$$
\Gamma_{1}^{-1}=\left(
\begin{array}{cc}
\left({\rm E}\left[HT\right]-{\rm E}\left[H\bld{X}^{\top}\right]{\rm E}\left[\bld{X}\bld{X}^{\top}\right]^{-1}{\rm E}\left[T\bld{X}\right]\right)^{-1}&-\frac{{\rm E}\left[H\bld{X}^{\top}\right]}{{\rm E}\left[HT\right]}\left({\rm E}\left[\bld{X}\bld{X}^{\top}\right]-\frac{{\rm E}\left[T\bld{X}\right]{\rm E}\left[H\bld{X}^{\top}\right]}{{\rm E}\left[HT\right]}\right)^{-1}\\
-\left({\rm E}\left[\bld{X}\bld{X}^{\top}\right]-\frac{{\rm E}\left[T\bld{X}\right]{\rm E}\left[H\bld{X}^{\top}\right]}{{\rm E}\left[HT\right]}\right)^{-1}\frac{{\rm E}\left[T\bld{X}\right]}{{\rm E}\left[HT\right]}&\left({\rm E}\left[\bld{X}\bld{X}^{\top}\right]-\frac{{\rm E}\left[T\bld{X}\right]{\rm E}\left[H\bld{X}^{\top}\right]}{{\rm E}\left[HT\right]}\right)^{-1}
\end{array}
\right).
$$
By using Sherman-Morrison formula (c.f. Harville, 2006), 
\begin{align*}
\left({\rm E}\left[\bld{X}\bld{X}^{\top}\right]-\frac{{\rm E}\left[T\bld{X}\right]{\rm E}\left[H\bld{X}^{\top}\right]}{{\rm E}\left[HT\right]}\right)^{-1}&\nonumber\\
&\hspace{-6.5cm}={\rm E}\left[\bld{X}\bld{X}^{\top}\right]^{-1}+\frac{{\rm E}\left[\bld{X}\bld{X}^{\top}\right]^{-1}{\rm E}\left[T\bld{X}\right]{\rm E}\left[H\bld{X}^{\top}\right]{\rm E}\left[\bld{X}\bld{X}^{\top}\right]^{-1}/{\rm E}\left[HT\right]}{1-{\rm E}\left[H\bld{X}^{\top}\right]{\rm E}\left[\bld{X}\bld{X}^{\top}\right]^{-1}{\rm E}\left[T\bld{X}\right]/{\rm E}\left[HT\right]}\nonumber\\
&\hspace{-6.5cm}=\scalebox{0.95}{$\displaystyle\frac{\left({\rm E}\left[HT\right]-{\rm E}\left[H\bld{X}^{\top}\right]{\rm E}\left[\bld{X}\bld{X}^{\top}\right]^{-1}{\rm E}\left[T\bld{X}\right]\right){\rm E}\left[\bld{X}\bld{X}^{\top}\right]^{-1}}{{\rm E}\left[HT\right]-{\rm E}\left[H\bld{X}^{\top}\right]{\rm E}\left[\bld{X}\bld{X}^{\top}\right]^{-1}{\rm E}\left[T\bld{X}\right]}+\frac{{\rm E}\left[\bld{X}\bld{X}^{\top}\right]^{-1}{\rm E}\left[T\bld{X}\right]{\rm E}\left[H\bld{X}^{\top}\right]{\rm E}\left[\bld{X}\bld{X}^{\top}\right]^{-1}}{{\rm E}\left[HT\right]-{\rm E}\left[H\bld{X}^{\top}\right]{\rm E}\left[\bld{X}\bld{X}^{\top}\right]^{-1}{\rm E}\left[T\bld{X}\right]}
$}
\end{align*}
Therefore,
\begin{align*}
\frac{{\rm E}\left[H\bld{X}^{\top}\right]}{{\rm E}\left[HT\right]}\left({\rm E}\left[\bld{X}\bld{X}^{\top}\right]-\frac{{\rm E}\left[T\bld{X}\right]{\rm E}\left[H\bld{X}^{\top}\right]}{{\rm E}\left[HT\right]}\right)^{-1}&=\frac{{\rm E}\left[H\bld{X}^{\top}\right]{\rm E}\left[\bld{X}\bld{X}^{\top}\right]^{-1}}{{\rm E}\left[HT\right]-{\rm E}\left[H\bld{X}^{\top}\right]{\rm E}\left[\bld{X}\bld{X}^{\top}\right]^{-1}{\rm E}\left[T\bld{X}\right]},\\
\left({\rm E}\left[\bld{X}\bld{X}^{\top}\right]-\frac{{\rm E}\left[T\bld{X}\right]{\rm E}\left[H\bld{X}^{\top}\right]}{{\rm E}\left[HT\right]}\right)^{-1}\frac{{\rm E}\left[T\bld{X}\right]}{{\rm E}\left[HT\right]}&=\frac{{\rm E}\left[\bld{X}\bld{X}^{\top}\right]^{-1}{\rm E}\left[T\bld{X}\right]}{{\rm E}\left[HT\right]-{\rm E}\left[H\bld{X}^{\top}\right]{\rm E}\left[\bld{X}\bld{X}^{\top}\right]^{-1}{\rm E}\left[T\bld{X}\right]}.
\end{align*}
We continue to calculate the variance components:
\begin{align}
\label{3_3}
\left(\Gamma_{1}\right)^{-1}\Sigma_{1}&=\sigma^2\left(
\begin{array}{cc}
\frac{{\rm E}\left[H^2\right]-{\rm E}\left[H\bld{X}^{\top}\right]{\rm E}\left[\bld{X}\bld{X}^{\top}\right]^{-1}{\rm E}\left[H\bld{X}\right]}{{\rm E}\left[HT\right]-{\rm E}\left[H\bld{X}^{\top}\right]{\rm E}\left[\bld{X}\bld{X}^{\top}\right]^{-1}{\rm E}\left[T\bld{X}\right]}&O^{\top}\\
\frac{{\rm E}\left[\bld{X}\bld{X}^{\top}\right]^{-1}\left({\rm E}[HT]{\rm E}\left[H\bld{X}\right]-{\rm E}\left[H^{2}\right]{\rm E}\left[T\bld{X}\right]\right)}{{\rm E}\left[HT\right]-{\rm E}\left[H\bld{X}^{\top}\right]{\rm E}\left[\bld{X}\bld{X}^{\top}\right]^{-1}{\rm E}\left[T\bld{X}\right]}&{\rm I}_{p}
\end{array}
\right),\nonumber\\
\left(\Gamma_{1}\right)^{-1}\Sigma_{1}\left(\Gamma_{1}^{\top}\right)^{-1}&=\sigma^2\left(
\begin{array}{cc}
\frac{{\rm E}\left[H^2\right]-{\rm E}\left[H\bld{X}^{\top}\right]{\rm E}\left[\bld{X}\bld{X}^{\top}\right]^{-1}{\rm E}\left[H\bld{X}\right]}{\left({\rm E}\left[HT\right]-{\rm E}\left[H\bld{X}^{\top}\right]{\rm E}\left[\bld{X}\bld{X}^{\top}\right]^{-1}{\rm E}\left[T\bld{X}\right]\right)^2}&\frac{\left({\rm E}\left[H^2\right]-{\rm E}\left[H\bld{X}^{\top}\right]{\rm E}\left[\bld{X}\bld{X}^{\top}\right]^{-1}{\rm E}\left[H\bld{X}\right]\right){\rm E}\left[T\bld{X}^{\top}\right]{\rm E}\left[\bld{X}\bld{X}^{\top}\right]^{-1}}{\left({\rm E}\left[HT\right]-{\rm E}\left[H\bld{X}^{\top}\right]{\rm E}\left[\bld{X}\bld{X}^{\top}\right]^{-1}{\rm E}\left[T\bld{X}\right]\right)^2}\\
&D
\end{array}
\right),
\end{align}
where
$$
D={\rm E}\left[\bld{X}\bld{X}^{\top}\right]^{-1}+\frac{{\rm E}\left[H^2\right]-{\rm E}\left[H\bld{X}^{\top}\right]{\rm E}\left[\bld{X}\bld{X}^{\top}\right]^{-1}{\rm E}\left[H\bld{X}\right]}{\left({\rm E}\left[HT\right]-{\rm E}\left[H\bld{X}^{\top}\right]{\rm E}\left[\bld{X}\bld{X}^{\top}\right]^{-1}{\rm E}\left[T\bld{X}\right]\right)^2}{\rm E}\left[\bld{X}\bld{X}^{\top}\right]^{-1}{\rm E}\left[T\bld{X}\right]{\rm E}\left[T\bld{X}^{\top}\right]{\rm E}\left[\bld{X}\bld{X}^{\top}\right]^{-1}.
$$
The (1,1) component of (\ref{3_3}) is
\begin{align}
\label{3_4}
\frac{{\rm E}\left[H^2\right]-{\rm E}\left[H\bld{X}^{\top}\right]{\rm E}\left[\bld{X}\bld{X}^{\top}\right]^{-1}{\rm E}\left[H\bld{X}\right]}{\left({\rm E}\left[HT\right]-{\rm E}\left[H\bld{X}^{\top}\right]{\rm E}\left[\bld{X}\bld{X}^{\top}\right]^{-1}{\rm E}\left[T\bld{X}\right]\right)^2}=\frac{\bld{\gamma}^{\top}\left({\rm E}\left[\boldsymbol{Z}^{\otimes 2}\right]-{\rm E}\left[\boldsymbol{Z}X^{\top}\right]{\rm E}\left[\bld{X}\bld{X}^{\top}\right]^{-1}{\rm E}\left[\bld{X}\boldsymbol{Z}^{\top}\right]\right)\bld{\gamma}}{\left(\bld{\gamma}^{\top}\left({\rm E}\left[\boldsymbol{Z}T\right]-{\rm E}\left[\boldsymbol{Z}\bld{X}^{\top}\right]{\rm E}\left[\bld{X}\bld{X}^{\top}\right]^{-1}{\rm E}\left[T\bld{X}\right]\right)\right)^2}.
\end{align}
By using {\bf Lemma \ref{ho3_1}.}, the minimum value of (\ref{3_4}) can be derived. Therefore, when
\begin{align}
\label{3_5}
\bld{\gamma}\propto \left({\rm E}\left[\boldsymbol{Z}^{\otimes 2}\right]-{\rm E}\left[\boldsymbol{Z}\bld{X}^{\top}\right]{\rm E}\left[\bld{X}\bld{X}^{\top}\right]^{-1}{\rm E}\left[\bld{X}\boldsymbol{Z}^{\top}\right]\right)^{-1}\left({\rm E}\left[\boldsymbol{Z}T\right]-{\rm E}\left[\boldsymbol{Z}\bld{X}^{\top}\right]{\rm E}\left[\bld{X}\bld{X}^{\top}\right]^{-1}{\rm E}\left[T\bld{X}\right]\right),
\end{align}
the asymptotic variance related to $\hat{\bld{\beta}}^{LC}$ becomes minimum.
\subsection{Proof of Theorem 1}\noindent
To describe simply, we use the following descriptions:
$$
\Sigma_{1}=\sigma^2\left(
\begin{array}{cc}
{\rm E}[H^{\otimes 2}]&{\rm E}[H\bld{X}^{\top}]\\
{\rm E}[\bld{X}H^{\top}]&{\rm E}[\bld{X}\bld{X}^{\top}]
\end{array}
\right)=\left(
\begin{array}{cc}
A&B^{\top}\\
B&C
\end{array}
\right),\ \Gamma_{1}=\left(
\begin{array}{cc}
{\rm E}\left[HT\right]&{\rm E}\left[H\bld{X}^{\top}\right]\\
{\rm E}\left[T\bld{X}\right]&{\rm E}\left[\bld{X}\bld{X}^{\top}\right]
\end{array}
\right)=\left(
\begin{array}{cc}
F&B^{\top}\\
G&C
\end{array}
\right)
$$
At first, we calculate the inverse matrix of $\Sigma_{1}$:
$$
\Sigma_{1}^{-1}=\left(
\begin{array}{cc}
\left(A-B^{\top}C^{-1}B\right)^{-1}&-A^{-1}B^{\top}\left(C-BA^{-1}B^{\top}\right)^{-1}\\
-\left(C-BA^{-1}B^{\top}\right)^{-1}BA^{-1}&\left(C-BA^{-1}B^{\top}\right)^{-1}
\end{array}
\right)=\left(
\begin{array}{cc}
D&-A^{-1}B^{\top}E\\
EBA^{-1}&E
\end{array}
\right),
$$
\begin{align}
\label{20200905_3}
\Gamma^{\top}_{1}\Sigma_{1}^{-1}&=\left(
\begin{array}{cc}
F^{\top}D-G^{\top}EBA^{-1}&-F^{\top}A^{-1}B^{\top}E+G^{\top}E\\
BD-CEBA^{-1}&-BA^{-1}B^{\top}E+CE
\end{array}
\right).
\end{align}
About the (2,2) component of (\ref{20200905_3}), by using 
$$
-BA^{-1}B^{\top}E+CE=\left(C-BA^{-1}B^{\top}\right)\left(C-BA^{-1}B^{\top}\right)^{-1}={\rm I}_{p},
$$
\begin{align}
\label{20200905_4}
\Gamma^{\top}_{1}\Sigma_{1}^{-1}\Gamma_{1}&\nonumber\\
&\hspace{-1.5cm}=\scalebox{0.9}{$\displaystyle\left(
\begin{array}{cc}
F^{\top}DF-G^{\top}EBA^{-1}F+G^{\top}EG-F^{\top}A^{-1}B^{\top}EG&F^{\top}DB^{\top}-G^{\top}EBA^{-1}B^{\top}+G^{\top}EC-F^{\top}A^{-1}B^{\top}EC\\
G+BDF-CEBA^{-1}F&BDB^{\top}-CEBA^{-1}B^{\top}+C
\end{array}
\right).
$}
\end{align}
From here, we calculate each component of (\ref{20200905_4}). At first, about (2,1):
\begin{align*}
BDF-CEBA^{-1}F&=\left(B-CEBA^{-1}\left(A-B^{\top}C^{-1}B\right)\right)DF\\
&=\left(B-CEB+CEBA^{-1}B^{\top}C^{-1}B\right)DF\\
&=\left(B-CE\left(C-BA^{-1}B^{\top}\right)C^{-1}B\right)DF\\
&=\left(B-C\left(C-BA^{-1}B^{\top}\right)^{-1}\left(C-BA^{-1}B^{\top}\right)C^{-1}B\right)DF\\
&=\bld{0}_{K}.
\end{align*}
Next, about (1,2):
\begin{align*}
-G^{\top}EBA^{-1}B^{\top}+G^{\top}EC&=G^{\top}E\left(C-BA^{-1}B^{\top}\right)\\
&=G^{\top}\left(C-BA^{-1}B^{\top}\right)^{-1}\left(C-BA^{-1}B^{\top}\right)\\
&=G^{\top}.
\end{align*}
At last, about (2,2):
\begin{align*}
BDB^{\top}-CEBA^{-1}B^{\top}&=O_{K}
\end{align*}
From the above, (\ref{20200905_4}) is
\begin{align}
\label{20200905_5}
\Gamma^{\top}_{1}\Sigma_{1}^{-1}\Gamma_{1}&=\left(
\begin{array}{cc}
F^{\top}DF-2G^{\top}EBA^{-1}F+G^{\top}EG&G^{\top}\\
G&C
\end{array}
\right),
\end{align}
Therefore, 
\begin{align}
\label{20200906_1}
\left(\Gamma^{\top}_{1}\Sigma_{1}^{-1}\Gamma_{1}\right)^{-1}&\nonumber\\
&\hspace{-2cm}=\scalebox{0.75}{$\left(
\begin{array}{cc}
\left(F^{\top}DF-2G^{\top}EBA^{-1}F+G^{\top}EG-G^{\top}C^{-1}G\right)^{-1}&-\left(F^{\top}DF-2G^{\top}EBA^{-1}F+G^{\top}EG-G^{\top}C^{-1}G\right)^{-1}G^{\top}C^{-1}\\
&C^{-1}+C^{-1}G\left(F^{\top}DF-2G^{\top}EBA^{-1}F+G^{\top}EG-G^{\top}C^{-1}G\right)^{-1}G^{\top}C^{-1}
\end{array}
\right)$}.
\end{align}
About the second term of (\ref{20200906_1}), 
\begin{align*}
G^{\top}EBA^{-1}F&=G^{\top}\left(C^{-1}+C^{-1}BDB^{\top}C^{-1}\right)BA^{-1}F\\
&=G^{\top}C^{-1}BA^{-1}F+G^{\top}C^{-1}BDB^{\top}C^{-1}BA^{-1}F\\
&=G^{\top}C^{-1}BD\left(\left(A-B^{\top}C^{-1}B\right)A^{-1}+B^{\top}C^{-1}BA^{-1}\right)F
&=G^{\top}C^{-1}BDF
\end{align*}
Next the third and fourth term of (\ref{20200906_1}),
\begin{align*}
G^{\top}EG-G^{\top}C^{-1}G&=G^{\top}C^{-1}G+G^{\top}C^{-1}BDB^{\top}C^{-1}G-G^{\top}C^{-1}G\\
&=G^{\top}C^{-1}BDB^{\top}C^{-1}G
\end{align*}
Therefore, (\ref{20200906_1}) becomes
\begin{align*}
\left(\Gamma^{\top}_{1}\Sigma_{1}^{-1}\Gamma_{1}\right)^{-1}_{(1,\, 1)}&=\left(FDF^{\top}-2G^{\top}C^{-1}BDF+G^{\top}C^{-1}BDB^{\top}C^{-1}G\right)^{-1}\\
&=\left(\left(F-B^{\top}C^{-1}G\right)^{\top}D\left(F-B^{\top}C^{-1}G\right)\right)^{-1}.
\end{align*}
Return to the original symbols:
\begin{align*}
\left(\Gamma^{\top}_{1}\Sigma_{1}^{-1}\Gamma_{1}\right)^{-1}_{(1,\, 1)}&\\
&\hspace{-2.5cm}=\sigma^{2}\left({\rm E}\left[HT\right]-{\rm E}\left[H\bld{X}^{\top}\right]{\rm E}\left[\bld{X}\bld{X}^{\top}\right]^{-1}{\rm E}\left[T\bld{X}\right]\right)^{\top}\left({\rm E}\left[H^{\otimes 2}\right]-{\rm E}\left[H\bld{X}^{\top}\right]{\rm E}\left[\bld{X}\bld{X}^{\top}\right]^{-1}{\rm E}\left[\bld{X}H^{\top}\right]\right)^{-1}\\
&\hspace{-2cm}\times\left({\rm E}\left[HT\right]-{\rm E}\left[H\bld{X}^{\top}\right]{\rm E}\left[\bld{X}\bld{X}^{\top}\right]^{-1}{\rm E}\left[T\bld{X}\right]\right)
\end{align*}
Therefore, the asymptotic variance related to $\hat{\beta}_{t}^{LC}$ consist with $\hat{\beta}_{t}^{GMM}$.  For the remaining components of (\ref{20200906_1}), the same discussion is considered.
\subsection{Proof of Lemma \ref{ho3_2}}\noindent
We proceed the proof by dividing (\ref{20200914_1}) into two parts: $\bld{c}_{i1},\, \bld{c}_{i2}(\bld{\gamma})$. At first, regarding $\bld{c}_{i1}$,
\begin{itemize}
\item When $k\in\{1,\dots,\ell\}$, 
\begin{align}
\label{20200914_2}
\frac{1}{n}\sum_{i=1}^{n}\left\{\left(Z_{ik}T_{i}-\sum_{j=1}^{p}\overline{Z_{k}X_{j}}\, \overline{TX_{j}}\right)I_{\tau k}+\kappa_{2n}\bar{I}_{\tau k}-\left(Z_{ik}T_{i}-\sum_{j=1}^{p}\overline{Z_{k}X_{j}}\, \overline{TX_{j}}\right)\right\}&\nonumber\\
&\hspace{-12cm}=\frac{1}{n}\sum_{i=1}^{n}\left\{\kappa_{2n}-\left(Z_{ik}T_{i}-\sum_{j=1}^{p}\overline{Z_{k}X_{j}}\, \overline{TX_{j}}\right)\right\}\bar{I}_{\tau k}
\end{align}
\item When $k\in\{\ell+1,\dots,K\}$, 
\begin{align}
\label{20200914_3}
\frac{1}{n}\sum_{i=1}^{n}\left\{\left(Z_{ik}T_{i}-\sum_{j=1}^{p}\overline{Z_{k}X_{j}}\, \overline{TX_{j}}\right)I_{\tau k}+\kappa_{2n}\bar{I}_{\tau k}-\kappa_{2n}\right\}&\nonumber\\
&\hspace{-8cm}=\frac{1}{n}\sum_{i=1}^{n}\left\{\left(Z_{ik}T_{i}-\sum_{j=1}^{p}\overline{Z_{k}X_{j}}\, \overline{TX_{j}}\right)-\kappa_{2n}\right\}I_{\tau k}
\end{align}
\end{itemize}
About (\ref{20200914_2}), (\ref{20200914_3}), since the first term ($\frac{1}{n}\sum_{i=1}^{n}\{\cdot\}$) convergence to a constant in probability, we need to confirm an asymptotic property of $I_{\tau k}$. Next, regarding $\bld{c}_{i2}(\bld{\gamma})$,
\begin{itemize}
\item When $k\in\{1,\dots,\ell\},\, k'\in\{1,\dots,\ell\}$,\\
About the diagonal component of the matrix,
\begin{align}
\label{20200914_4}
\frac{1}{n}\sum_{i=1}^{n}\left\{\left(Z_{ik}^2-\sum_{j=1}^{p}\overline{Z_{k}X_{j}}^2\right)I_{\tau k}+\kappa_{1}\bar{I}_{\tau k}-\left(Z_{ik}^2-\sum_{j=1}^{p}\overline{Z_{k}X_{j}}^2\right)\right\}&\nonumber\\
&\hspace{-8cm}=\frac{1}{n}\sum_{i=1}^{n}\left\{\kappa_{1}-\left(Z_{ik}^2-\sum_{j=1}^{p}\overline{Z_{k}X_{j}}^2\right)\right\}\bar{I}_{\tau k}
\end{align}
About the non-diagonal component of the matrix,
\begin{align}
\label{20200914_5}
\frac{1}{n}\sum_{i=1}^{n}\left\{\left(Z_{ik}Z_{ik'}-\sum_{j=1}^{p}\overline{Z_{k}X_{j}}\, \overline{Z_{k'}X_{j}}\right)I_{\tau k}I_{\tau k'}-\left(Z_{ik}Z_{ik'}-\sum_{j=1}^{p}\overline{Z_{k}X_{j}}\, \overline{Z_{k'}X_{j}}\right)\right\}\nonumber\\
&\hspace{-12cm}=\frac{1}{n}\sum_{i=1}^{n}\left\{\left(Z_{ik}Z_{ik'}-\sum_{j=1}^{p}\overline{Z_{k}X_{j}}\, \overline{Z_{k'}X_{j}}\right)\right\}\left(I_{\tau k}I_{\tau k'}-1\right)
\end{align}
\item When $k\in\{\ell+1,\dots,K\},\, k'\in\{\ell+1,\dots,K\}$,\\
About the diagonal component of the matrix,
\begin{align}
\label{20200914_6}
\frac{1}{n}\sum_{i=1}^{n}\left\{\left(Z_{ik}T_{i}-\sum_{j=1}^{p}\overline{Z_{k}X_{j}}\, \overline{TX_{j}}\right)I_{\tau k}+\kappa_{2n}\bar{I}_{\tau k}-\kappa_{2n}\right\}&\nonumber\\
&\hspace{-8cm}=\frac{1}{n}\sum_{i=1}^{n}\left\{\left(Z_{ik}T_{i}-\sum_{j=1}^{p}\overline{Z_{k}X_{j}}\, \overline{TX_{j}}\right)-\kappa_{2n}\right\}I_{\tau k}
\end{align}
\item When $k\in\{1,\dots,K\},\, k'\in\{\ell+1,\dots,K\}$, or $k\in\{\ell+1,\dots,K\},\, k'\in\{1,\dots,K\}$,\\
About the non-diagonal component of the matrix,
\begin{align}
\label{20200914_7}
\frac{1}{n}\sum_{i=1}^{n}\left(Z_{ik}Z_{ik'}-\sum_{j=1}^{p}\overline{Z_{k}X_{j}}\, \overline{Z_{k'}X_{j}}\right)I_{\tau k}I_{\tau k'}
\end{align}
\end{itemize}
About (\ref{20200914_4})-(\ref{20200914_7}), since the first term ($\frac{1}{n}\sum_{i=1}^{n}\{\cdot\}$) convergence to a constant in probability, we need to confirm an asymptotic property of $I_{\tau k}$.

From here, we confirm when $I_{\tau k}=o_{p}\left(1/\sqrt{n}\right)$ or $\bar{I}_{\tau k}=o_{p}\left(1/\sqrt{n}\right)$ hold. Note that under this situation, the product ($I_{\tau k}I_{\tau k'}$) becomes $o_{p}\left(1/n\right)$. Regarding $I_{\tau k}$,
$$
I_{\tau k}=\Phi_{\tau}(\hat{w}_{k}+w)\Phi_{\tau}(w-\hat{w}_{k})=\frac{1}{\sqrt{2\pi\tau^2}}\int_{-\infty}^{\hat{w}_{k}+w}\exp\left\{-\frac{\omega^2}{2\tau^2}\right\}d\omega\frac{1}{\sqrt{2\pi\tau^2}}\int_{-\infty}^{w-\hat{w}_{k}}\exp\left\{-\frac{\omega^2}{2\tau^2}\right\}d\omega.
$$
Transformation of variables as $\omega'=\omega/\tau$, 
$$
I_{\tau k}=\frac{1}{\sqrt{2\pi}}\int_{-\infty}^{\frac{\hat{w}_{k}+w}{\tau}}\exp\left\{-\frac{(\omega')^2}{2}\right\}d\omega'\frac{1}{\sqrt{2\pi}}\int_{-\infty}^{\frac{w-\hat{w}_{k}}{\tau}}\exp\left\{-\frac{(\omega')^2}{2}\right\}d\omega'=\int_{-\infty}^{\frac{\hat{w}_{k}+w}{\tau}}d\Phi\int_{-\infty}^{\frac{w-\hat{w}_{k}}{\tau}}d\Phi,
$$
where $\Phi$ is CDF of $N(0,1)$. Then
\begin{align}
\label{20200914_10}
0<1-I_{\tau k}&=1-\int_{-\infty}^{\frac{\hat{w}_{k}+w}{\tau}}d\Phi\int_{-\infty}^{\frac{w-\hat{w}_{k}}{\tau}}d\Phi\nonumber\\
&< \left\{
\begin{array}{c}
1-\left(\int_{-\infty}^{\frac{w-\hat{w}_{k}}{\tau}}d\Phi\right)^2\ \ \ (\hat{w}_{k}\geq 0)\\
1-\left(\int_{-\infty}^{\frac{\hat{w}_{k}+w}{\tau}}d\Phi\right)^2\ \ \ (\hat{w}_{k}< 0)
\end{array}
\right.\nonumber\\
&=1-\left(\int_{-\infty}^{\frac{w-|\hat{w}_{k}|}{\tau}}d\Phi\right)^2,
\end{align}
and
\begin{align}
\label{20200914_11}
0<I_{\tau k}=\int_{-\infty}^{\frac{\hat{w}_{k}+w}{\tau}}d\Phi\int_{-\infty}^{\frac{w-\hat{w}_{k}}{\tau}}d\Phi<\int_{-\infty}^{\frac{w-|\hat{w}_{k}|}{\tau}}d\Phi
\end{align}
satisfy.
\begin{itemize}
\item When $k\in\{1,\dots,\ell\}$,\\
In this situation, $\hat{w}_{k}\stackrel{P}{\to}0$. At first, when $w\geq |\hat{w}_{k}|$, by using (\ref{20200914_10})
\begin{align}
\label{20200914_8}
0<1-I_{\tau_{n} k}&=1-\left(\int_{-\infty}^{\frac{w-|\hat{w}_{k}|}{\tau_{n}}}d\Phi\right)^2=1-\left(1-\int_{\frac{w-|\hat{w}_{k}|}{\tau_{n}}}^{\infty}d\Phi\right)^2\nonumber\\
&=2\int_{\frac{w-|\hat{w}_{k}|}{\tau_{n}}}^{\infty}d\Phi-\left(\int_{\frac{w-|\hat{w}_{k}|}{\tau_{n}}}^{\infty}d\Phi\right)^2<2\int_{\frac{w-|\hat{w}_{k}|}{\tau_{n}}}^{\infty}d\Phi
\end{align}
Regarding (\ref{20200914_8}), using an evaluation of the tail of Normal distribution (c.f. Gordon, 1941),
\begin{align*}
0<1-I_{\tau_{n} k}&<2\int_{\frac{w-|\hat{w}_{k}|}{\tau_{n}}}^{\infty}d\Phi-\left(\int_{\frac{w-|\hat{w}_{k}|}{\tau_{n}}}^{\infty}d\Phi\right)^2<2\int_{\frac{w-|\hat{w}_{k}|}{\tau_{n}}}^{\infty}d\Phi\\
&<\frac{2}{\sqrt{2\pi}}\left(\frac{w-|\hat{w}_{k}|}{\tau_{n}}\right)^{-1}\exp\left\{-\frac{1}{2}\frac{\left(w-|\hat{w}_{k}|\right)^2}{\tau_{n}^2}\right\}\\
&<\left(\frac{|w-\hat{w}_{k}|}{\tau_{n}}\right)^{-1}\exp\left\{-\frac{1}{2}\frac{\left(w-\hat{w}_{k}\right)^2}{\tau_{n}^2}\right\}.
\end{align*}
Therefore, when
\begin{align}
\label{20200914_9}
\left(\frac{|w-\hat{w}_{k}|}{\tau_{n}}\right)^{-1}\exp\left\{-\frac{1}{2}\frac{\left(w-\hat{w}_{k}\right)^2}{\tau_{n}^2}\right\}=o_{p}\left(\frac{1}{\sqrt{n}}\right),
\end{align}
$\bar{I}_{\tau k}=o_{p}\left(1/\sqrt{n}\right)$ is satisfied, and $I_{\tau k}=1+o_{p}\left(1/\sqrt{n}\right)$ is also satisfied. Next, when $w<|\hat{w}_{k}|$, we can ignore the situation in the view of the convergence in probability since $\hat{w}_{k}\stackrel{P}{\to}0$.
\item When $k\in\{\ell+1,\dots,K\}$,\\
In this situation, $\hat{w}_{k}\stackrel{P}{\to}w_{k}$. At first, when $w\geq |\hat{w}_{k}|$, we can ignore the situation in the view of the convergence in probability since $|\hat{w}_{k}|\stackrel{P}{\to}|w_{k}|>w$. Next, when $w<|\hat{w}_{k}|$, by using (\ref{20200914_11}) and an evaluation of the tail of Normal distribution,
\begin{align*}
0<I_{\tau_{n} k}&<\int_{-\infty}^{\frac{w-|\hat{w}_{k}|}{\tau}}d\Phi=\int^{\infty}_{-\frac{w-|\hat{w}_{k}|}{\tau}}d\Phi<\left(\frac{|w-\hat{w}_{k}|}{\tau_{n}}\right)^{-1}\exp\left\{-\frac{1}{2}\frac{\left(w-\hat{w}_{k}\right)^2}{\tau_{n}^2}\right\}.
\end{align*}
Therefore, when (\ref{20200914_9}), $I_{\tau k}=o_{p}\left(1/\sqrt{n}\right)$ hold, and $\bar{I}_{\tau k}=1+o_{p}\left(1/\sqrt{n}\right)$ also hold.
\end{itemize}
From the above, when (\ref{20200914_9}) is satisfied, (\ref{20200914_1}) is also satisfied.
\subsection{Proof of Proposition \ref{te3_1}}\noindent
From the result of {\bf Lemma \ref{ho3_2}.}, $\forall\bld{\gamma}$,
$$
\frac{1}{n}\sum_{i=1}^{n}\left(
\bld{b}_{i}-A_{i}\bld{\gamma}
\right)-\frac{1}{n}\sum_{i=1}^{n}\left\{
\left(
\begin{array}{c}
\check{\bld{b}}_{i}\\
\kappa_{2n}\bld{1}_{K-\ell}
\end{array}
\right)-\left(
\begin{array}{cc}
\check{A}_{i}&O_{\ell\times K-\ell}\\
O_{\ell\times K-\ell}^{\top}&\kappa_{1}{\rm I}_{K-\ell}
\end{array}
\right)\bld{\gamma}
\right\}=o_{p}\left(\frac{1}{\sqrt{n}}\right).
$$
Therefore,
\begin{align}
\label{20200913_1}
\bld{0}_{K}=\frac{1}{n}\sum_{i=1}^{n}\left(
\bld{b}_{i}-A_{i}\hat{\bld{\gamma}}
\right)
=\frac{1}{n}\sum_{i=1}^{n}\left\{
\left(
\begin{array}{c}
\check{\bld{b}}_{i}\\
\kappa_{2n}\bld{1}_{K-\ell}
\end{array}
\right)-\left(
\begin{array}{cc}
\check{A}_{i}&O_{\ell\times K-\ell}\\
O_{\ell\times K-\ell}^{\top}&\kappa_{1}{\rm I}_{K-\ell}
\end{array}
\right)\hat{\bld{\gamma}}
\right\}+o_{p}\left(\frac{1}{\sqrt{n}}\right),
\end{align}
\begin{align}
\label{for_tun_1}
\hat{\bld{\gamma}}=\left(
\frac{1}{n}\sum_{i=1}^{n}\left(
\begin{array}{cc}
\check{A}_{i}&O_{\ell\times K-\ell}\\
O_{\ell\times K-\ell}^{\top}&\kappa_{1}{\rm I}_{K-\ell}
\end{array}
\right)
\right)^{-1}\frac{1}{n}\sum_{i=1}^{n}\left(
\begin{array}{c}
\check{\bld{b}}_{i}\\
\kappa_{2n}\bld{1}_{K-\ell}
\end{array}
\right)+o_{p}\left(\frac{1}{\sqrt{n}}\right)\stackrel{P}{\to}\left(
\begin{array}{c}
{\rm E}\left[\check{A}\right]^{-1}{\rm E}\left[\check{\bld{b}}\right]\\
\bld{0}_{K-\ell}
\end{array}
\right)
\end{align}
From the above the first component of {\bf Proposition \ref{te3_1}} can be proved. Next regarding (\ref{20200913_1}), conducting the taylor expansion around $\bld{\gamma}^{0}$,
\begin{align*}
\bld{0}_{K}&=\frac{1}{n}\sum_{i=1}^{n}\left\{
\left(
\begin{array}{c}
\check{\bld{b}}_{i}\\
\kappa_{2n}\bld{1}_{K-\ell}
\end{array}
\right)-\left(
\begin{array}{cc}
\check{A}_{i}&O_{\ell\times K-\ell}\\
O_{\ell\times K-\ell}^{\top}&\kappa_{1}{\rm I}_{K-\ell}
\end{array}
\right)\hat{\bld{\gamma}}
\right\}+o_{p}\left(\frac{1}{\sqrt{n}}\right)\\
&=\frac{1}{n}\sum_{i=1}^{n}\left\{
\left(
\begin{array}{c}
\check{\bld{b}}_{i}\\
\kappa_{2n}\bld{1}_{K-\ell}
\end{array}
\right)-\left(
\begin{array}{cc}
\check{A}_{i}&O_{\ell\times K-\ell}\\
O_{\ell\times K-\ell}^{\top}&\kappa_{1}{\rm I}_{K-\ell}
\end{array}
\right)\bld{\gamma}^{0}
\right\}\\
&\hspace{0.5cm}-\frac{1}{n}\sum_{i=1}^{n}\left(
\begin{array}{cc}
\check{A}_{i}&O_{\ell\times K-\ell}\\
O_{\ell\times K-\ell}^{\top}&\kappa_{1}{\rm I}_{K-\ell}
\end{array}
\right)\left(\hat{\bld{\gamma}}-\bld{\gamma}^{0}\right)+o_{p}\left(\frac{1}{\sqrt{n}}\right).
\end{align*}
Therefore, 
\begin{align*}
\sqrt{n}\left(\hat{\bld{\gamma}}-\bld{\gamma}^{0}\right)&=\left(\frac{1}{n}\sum_{i=1}^{n}\left(
\begin{array}{cc}
\check{A}_{i}&O_{\ell\times K-\ell}\\
O_{\ell\times K-\ell}^{\top}&\kappa_{1}{\rm I}_{K-\ell}
\end{array}
\right)\right)^{-1}\\
&\hspace{0.5cm}\times\frac{\sqrt{n}}{n}\sum_{i=1}^{n}\left\{
\left(
\begin{array}{c}
\check{\bld{b}}_{i}\\
\kappa_{2n}\bld{1}_{K-\ell}
\end{array}
\right)-\left(
\begin{array}{cc}
\check{A}_{i}&O_{\ell\times K-\ell}\\
O_{\ell\times K-\ell}^{\top}&\kappa_{1}{\rm I}_{K-\ell}
\end{array}
\right)\bld{\gamma}^{0}
\right\}+o_{p}(1)\\
&=\left(
\begin{array}{cc}
\left(\frac{1}{n}\sum_{i=1}^{n}\check{A}_{i}\right)^{-1}&O_{\ell\times K-\ell}\\
O_{\ell\times K-\ell}^{\top}&\frac{1}{\kappa_{1}}{\rm I}_{K-\ell}
\end{array}
\right)
\left(
\begin{array}{c}
\frac{\sqrt{n}}{n}\sum_{i=1}^{n}\left(\check{\bld{b}}_{i}-\check{A}_{i}\bld{\gamma}^{0}_{val}\right)\\
\sqrt{n}\kappa_{2n}\bld{1}_{K-\ell}
\end{array}
\right)
+o_{p}(1)\\
&=
\left(
\begin{array}{c}
\left(\frac{1}{n}\sum_{i=1}^{n}\check{A}_{i}\right)^{-1}\frac{\sqrt{n}}{n}\sum_{i=1}^{n}\left(\check{\bld{b}}_{i}-\check{A}_{i}\bld{\gamma}^{0}_{val}\right)\\
\sqrt{n}\frac{\kappa_{2n}}{\kappa_{1}}\bld{1}_{K-\ell}
\end{array}
\right)
+o_{p}(1)
\end{align*}
If $\left|{\rm E}\left[\left(\check{\bld{b}}-\check{A}\bld{\gamma}_{val}^{0}\right)^{\otimes 2}\right]\right|<\infty$, by using the ordinary asymptotic theories for M-estimator (c.f. Van der Vaart, 2000), the second component of {\bf Proposition \ref{te3_1}} can be proved. 
\subsection{Proof of Theorem 2}\noindent
Regarding (3.6) in the main manuscript, conducting the taylor expansion around $\bld{\beta}^{0}$ and $\bld{\gamma}^{0}$,
\begin{align*}
\bld{0}_{p+1}&=\frac{1}{n}\sum_{i=1}^{n}\left(
\begin{array}{c}
\hat{\bld{\gamma}}^{\top}\bld{z}_{i}\\
\bld{x}_{i}
\end{array}
\right)\left(y_{i}-(t_{i},\, \bld{x}_{i}^{\top})\hat{\boldsymbol{\beta}}\right)\\
&=\frac{1}{n}\sum_{i=1}^{n}\left(
\begin{array}{c}
\left(\bld{\gamma}^{0}\right)^{\top}\bld{z}_{i}\\
\bld{x}_{i}
\end{array}
\right)\left(y_{i}-(t_{i},\, \bld{x}_{i}^{\top})\boldsymbol{\beta}^{0}\right)-\frac{1}{n}\sum_{i=1}^{n}\left(
\begin{array}{cc}
\left(\bld{\gamma}^{0}\right)^{\top}\bld{z}_{i}t_{i}&\left(\bld{\gamma}^{0}\right)^{\top}\bld{z}_{i}\bld{x}_{i}^{\top}\\
\bld{x}_{i}t_{i}&\bld{x}_{i}\bld{x}_{i}^{\top}
\end{array}
\right)\left(\hat{\boldsymbol{\beta}}-\boldsymbol{\beta}^{0}\right)\\
&\hspace{0.5cm}+\frac{1}{n}\sum_{i=1}^{n}\left(
\begin{array}{c}
\left(y_{i}-(t_{i},\, \bld{x}_{i}^{\top})\boldsymbol{\beta}^{0}\right)\bld{z}_{i}^{\top}\\
O_{p\times K}
\end{array}
\right)\left(\hat{\boldsymbol{\gamma}}-\boldsymbol{\gamma}^{0}\right)+o_{p}\left(\frac{1}{\sqrt{n}}\right).
\end{align*}
Therefore,
\begin{align}
\label{20200913_2}
\sqrt{n}\left(\hat{\boldsymbol{\beta}}-\boldsymbol{\beta}^{0}\right)&=\left(
\frac{1}{n}\sum_{i=1}^{n}\left(
\begin{array}{cc}
\left(\bld{\gamma}^{0}\right)^{\top}\bld{z}_{i}t_{i}&\left(\bld{\gamma}^{0}\right)^{\top}\bld{z}_{i}\bld{x}_{i}^{\top}\\
\bld{x}_{i}t_{i}&\bld{x}_{i}\bld{x}_{i}^{\top}
\end{array}
\right)
\right)^{-1}\left\{
\frac{\sqrt{n}}{n}\sum_{i=1}^{n}\left(
\begin{array}{c}
\left(\bld{\gamma}^{0}\right)^{\top}\bld{z}_{i}u_{i}\\
\bld{x}_{i}u_{i}
\end{array}
\right)\right.\nonumber\\
&\hspace{0.5cm}\left.+\frac{1}{n}\sum_{i=1}^{n}\left(
\begin{array}{c}
u_{i}\bld{z}_{i}^{\top}\\
O_{p\times K}
\end{array}
\right)\sqrt{n}\left(\hat{\boldsymbol{\gamma}}-\boldsymbol{\gamma}^{0}\right)
\right\}+o_{p}\left(\frac{1}{\sqrt{n}}\right).
\end{align}
The second term of $\{\cdot\}$ in (\ref{20200913_2}) becomes 
$$
\frac{1}{n}\sum_{i=1}^{n}u_{i}\bld{z}_{i}\stackrel{P}{\to}{\rm E}\left[UZ\right]=\left(
\begin{array}{c}
\bld{0}_{\ell}\\
{\rm E}\left[UZ_{\ell+1}\right]\\
\vdots\\
{\rm E}\left[UZ_{K}\right]
\end{array}
\right),
$$
and the result of {\bf Proposition \ref{te3_1}.}, we can show that  
$$
\frac{1}{n}\sum_{i=1}^{n}\left(
\begin{array}{c}
u_{i}\bld{z}_{i}^{\top}\\
O_{p\times K}
\end{array}
\right)\sqrt{n}\left(\hat{\boldsymbol{\gamma}}-\boldsymbol{\gamma}^{0}\right)=o_{p}(1).
$$
Therefore, by using the ordinary asymptotic theories for M-estimator to (\ref{20200913_2}), {\bf Theorem 2} can be proved.
\section{Using valid IVs as auxiliary variables}
\label{App2}
Assume that we know at least one IV $Z_{0}$ is valid. Therefore, a solution of the following estimating equation becomes true causal effects when (2.1) is the true outcome model:
$$
\sum_{i=1}^{n}\left(
\begin{array}{c}
z_{0i}\\
\bld{x}_{i}
\end{array}
\right)\left(y_{i}-(t_{i},\bld{x}^{\top}_{i})\boldsymbol{\beta}\right)=\bld{0}_{p+1}
$$
Also, residuals can be estimated as follows:
$$
\varepsilon_{i}\left(\hat{\bld{\beta}}_{0}\right)=y_{i}-(t_{i},\bld{x}^{\top}_{i})\hat{\bld{\beta}}_{0}
$$
When constructing $\hat{w}_{k}$, we substitute $\varepsilon_{i}\left(\hat{\bld{\beta}}_{0}\right)$ for NCO $M_{i}$:
$$
\hat{w}_{k}=\frac{1}{n}\sum_{i=1}^{n}Z_{k i}\varepsilon_{i}\left(\hat{\bld{\beta}}_{0}\right),
$$
and
$$
\hat{w}_{k}\stackrel{P}{\to}{\rm E}\left[Z_{k}U\right].
$$
And, we can easily confirm the discussions and proofs of {\bf Section 3}. Therefore, we can identify valid and invalid IVs by using $Z_{0}$ as auxiliary variables.

\section{How to handle tuning parameters}
To implement our proposed method, we have to decide tuning parameters $(\kappa_{1},\kappa_{2n},\tau,w)$. $\kappa_{1}$ and $\kappa_{2n}$ control the variability of $\hat{\bld{\gamma}}_{inv}$ (see (\ref{for_tun_1})). As is clear from (\ref{for_tun_1}), $\kappa_{1}$ should be set as large as possible and $\kappa_{2n}$ should be set as small as possible. From these settings, we can estimate $\hat{\bld{\gamma}}_{inv}$ near $0$. $\tau$ is also easy to be set since it is only use for the proof of Theorem 2. Therefore, to implement the proposed method, $\tau$ should be set as small as possible.

However, $w$ is hard to decide and need to be decided very carefully. This is because $w$ decide the cut-off point whether the candidates of IVs are valid or not. When $w$ is set as small value, invalid IVs tend not to be selected; whereas, some valid IVs may not be selected. Therefore, the efficiency of the estimator becomes decrease. On the other hand, when $w$ is set as large value, the inefficiency is improved; whereas, some invalid IVs may not be selected. Therefore, the estimator may have some bias. This relationship is similar as type 1 error and type 2 error in the context of statistical tests. To decide the cut-off point, we consider a decision method following statistical test contexts. We consider the statistics (3.4). Since the statistics is an ordinary sample mean, it becomes
$$
\frac{\frac{1}{\sqrt{n}}\sum_{i=1}^{n}Z_{k i}M_{i}}{\sqrt{{\rm E}[Z_{k}^2]{\rm E}[M^2]}}\stackrel{L}{\to}N(0,1),
$$
when $Z_{k}$ is a valid IV. Then, we consider the probability $\alpha$ that is ``at least one covariance between a valid IV and the NCO exceeds a cut-off point $w$". Under these settings, $w$ can be decided to satisfy the following probability:
$$
{\rm Pr}\left(\frac{\frac{1}{\sqrt{n}}\sum_{i=1}^{n}|Z_{k i}M_{i}|}{\sqrt{{\rm E}[Z_{k}^2]{\rm E}[M^2]}}>w\right)\leq \frac{\alpha}{K}
$$
Actually,
$$
{\rm Pr}\left(\bigcup_{k=1}^{K}\left\{\frac{\frac{1}{\sqrt{n}}\sum_{i=1}^{n}|Z_{k i}M_{i}|}{\sqrt{{\rm E}[Z_{k}^2]{\rm E}[M^2]}}>w\right\}\right)\leq\sum_{k=1}^{K}{\rm Pr}\left(\frac{\frac{1}{\sqrt{n}}\sum_{i=1}^{n}|Z_{k i}M_{i}|}{\sqrt{{\rm E}[Z_{k}^2]{\rm E}[M^2]}}>w\right)\leq\alpha.
$$
Specifically, $w$ becomes $|z_{\frac{\alpha}{2K}}|$, where $z_{\alpha}$ is the $\alpha$ percent point of the standard normal distribution. We use the cut-off point for the following simulations and data analyses.

On the other hand, we can also consider the probability $1-\beta$ that is ``all covariances between invalid IVs and the NCO exceed the cut-off point $w$". When $Z_{k}$ is an invalid IV,
$$
\frac{\frac{1}{\sqrt{n}}\sum_{i=1}^{n}Z_{k i}M_{i}}{\sqrt{{\rm E}[Z_{k}^2]{\rm E}[M^2]}}\stackrel{L}{\to}N\left(\sqrt{n}\frac{{\rm E}[Z_{k}M]}{\sqrt{{\rm E}[Z_{k}^2]{\rm E}[M^2]}},\frac{Var(Z_{k}M)}{{\rm E}[Z_{k}^2]{\rm E}[M^2]}\right)
$$
from the similar calculation as previous one. Then,
\begin{align*}
1-\beta&={\rm Pr}\left(\bigcap_{k=\ell+1}^{K}\left\{\left\{\frac{\frac{1}{\sqrt{n}}\sum_{i=1}^{n}Z_{k i}M_{i}}{\sqrt{{\rm E}[Z_{k}^2]{\rm E}[M^2]}}<-w\right\}\cup\left\{\frac{\frac{1}{\sqrt{n}}\sum_{i=1}^{n}Z_{k i}M_{i}}{\sqrt{{\rm E}[Z_{k}^2]{\rm E}[M^2]}}>w\right\}\right\}\right)\\
&\geq \sum_{k=\ell+1}^{K}{\rm Pr}\left(\left\{\left\{\frac{\frac{1}{\sqrt{n}}\sum_{i=1}^{n}Z_{k i}M_{i}}{\sqrt{{\rm E}[Z_{k}^2]{\rm E}[M^2]}}<-w\right\}\cup\left\{\frac{\frac{1}{\sqrt{n}}\sum_{i=1}^{n}Z_{k i}M_{i}}{\sqrt{{\rm E}[Z_{k}^2]{\rm E}[M^2]}}>w\right\}\right\}\right)-(K-\ell-1)\\
&=1-\sum_{k=\ell+1}^{K}{\rm Pr}\left(\frac{\frac{1}{\sqrt{n}}\sum_{i=1}^{n}|Z_{k i}M_{i}|}{\sqrt{{\rm E}[Z_{k}^2]{\rm E}[M^2]}}<w\right)\\
&\geq1-\sum_{k=\ell+1}^{K}\Phi\left(\frac{w-\sqrt{n}\frac{{\rm E}[Z_{k}M]}{\sqrt{{\rm E}[Z_{k}^2]{\rm E}[M^2]}}}{\sqrt{\frac{Var(Z_{k}M)}{{\rm E}[Z_{k}^2]{\rm E}[M^2]}}}\right).
\end{align*}
Therefore, when $n\to \infty$, the above probability becomes sufficient large; whereas, when $K-\ell$ is relatively large, the above probability may not become sufficient large. Also, the probability depends on the correlation between invalid IVs and the NCO: ${\rm E}[Z_{k}M]$. In short, the power ``$1-\beta$" depends on complex effects.

\section{Supplementary information of simulations}
\subsection{Setting of Liao, 2013}
In this simulation, our proposed method is compared with the previous methods: the ordinary GMM and proposed by Liao, 2013 and DiTraglia, 2016. The simulation setting is as follows:
\begin{description}
\item{{\bf Candidates of IVs, treatment, and unmeasured covariate}}
$$
\left(T_{i},\, Z_{i1},\, Z_{i21},\, U_{i},\, Z_{i22}^{*}\right)^{\top}\stackrel{i.i.d.}{\sim}N_{5}\left(\bld{0}_{5},\, \Sigma\right),\ Z_{i22}=Z_{i22}^{*}+0.5*U_{i}
$$
where
$$
\Sigma=\left(
\begin{array}{ccccc}
1&\sigma_{tz_{1}}&\sigma_{tz_{21}}&0.4&0\\
&1&0&0&0\\
&&1&0&0\\
&&&1&0\\
&&&&1
\end{array}
\right)
$$
\begin{enumerate}
\item Strong IVs: $(\sigma_{tz_{1}},\sigma_{tz_{21}})=(0.4,0.4)$
\item Weak IVs: $(\sigma_{tz_{1}},\sigma_{tz_{21}})=(0.1,0.3)$
\end{enumerate}
\item{{\bf Outcome}}
$$
Y_{i}=0.8+0.8T_{i}+U_{i}
$$
\item{{\bf Negative Control Outcome}}
$$
M_{i}=1+\alpha_{mu}U_{i}+\varepsilon_{mi},\ \ \varepsilon_{mi}\sim N(0,1)
$$
\begin{enumerate}
\item Strong NCO: $\alpha_{mu}=1$\\
$\Rightarrow$Correlation between a NCO and an unmeasured covariate becomes approximately 0.71.
\item Weak NCO: $\alpha_{mu}=0.35$\\
$\Rightarrow$Correlation between a NCO and an unmeasured covariate becomes approximately 0.34.
\end{enumerate}
\end{description}
Note that $Z_{22}$ is the invalid IV in the candidates of IVs; we would like to use $Z_{1},\, Z_{21}$ to estimate an unbiased causal effect. In this simulation, we confirm the two scenarios:
\begin{enumerate}
\item We know some valid IVs (i.e. $Z_{1}$) and NCO (i.e. $M$)
\item We do not know valid IVs, but know NCO
\end{enumerate}
Therefore, in scenario 1, all methods work well; however, in scenario 2, the previous method do not work whereas our proposed method works. Regarding the method of Liao, 2013, we use an Adaptive LASSO-type penalty and tuning parameters are selected as $\lambda_{n}=0.5$ and $\omega=1$. Regarding our proposed method, the tuning parameters are selected as $(\kappa_{1},\kappa_{2n},\tau)=(10^{8},0.001,10^{-6})$. Also, the important tuning parameter $w$ is selected as described in section 3.2 with $\alpha=0.1$ (i.e. the probability of ``at least one covariance between a valid IV and the NCO exceeds a cut-off point w" is 0.1).

The simulation results are summarized in tables and supplemental figures. The estimated causal effects are summarized in Table \ref{tab21}, Figure \ref{fig1}, \ref{fig2}, \ref{fig3}, and \ref{fig4}. Our proposed method has the similar efficiency as GMM when there is the strong IV and strong NCO. Whereas, our proposed method has larger variance than GMM when there is the strong IV and weak NCO. Therefore, our proposed method is affected by the strength of the available NCO. This features are similar as weak IV situation. Regarding the methods of Liao, 2013 and DiTraglia, 2016, their methods have also good performance both the strong IV and the weak IV situations when the valid IVs are specified. However, when the valid IVs are unspecified, these methods have obvious bias; detecting the valid IVs before analyses is important role for their methods.

\begin{landscape}
\begin{table}[h]
\begin{center}
\caption{Summary of estimators for causal effects by situations}
\scalebox{0.85}{
\begin{tabular}{|c|c|c|c||c|c|c|c||c|c|c|c|}\hline
{\bf Information}&{\bf Setting}&{\bf Setting}&{\bf Method}&\multicolumn{4}{|c||}{Small sample $N=500$}&\multicolumn{4}{|c|}{Large sample $N=1000$}\\\cline{5-12}
{\bf of IV}&{\bf of IV}&{\bf of NCO}&&Mean (SD)&Median (Range)&$|$Bias$|$&RMSE&Mean (SD)&Median (Range)&$|$Bias$|$&RMSE\\\hline
-&\begin{tabular}{c}Strong\\IV\end{tabular}&\begin{tabular}{c}Strong\\NCO\end{tabular}&{\bf Proposed}&0.801 (0.083)&0.802 (0.50, 1.33)&0.001&0.083&0.798 (0.054)&0.801 (0.61, 0.97)&0.002&0.054\\\cline{3-12}
&&\begin{tabular}{c}Weak\\NCO\end{tabular}&{\bf Proposed}&0.796 (0.247)&0.810 (-4.57, 2.92)&0.004&0.247&0.806 (0.111)&0.810 (0.10, 1.42)&0.006&0.111\\\cline{2-12}
&\begin{tabular}{c}Weak\\IV\end{tabular}&\begin{tabular}{c}Strong\\NCO\end{tabular}&{\bf Proposed}&0.812 (0.099)&0.807 (0.38, 1.39)&0.012&0.100&0.799 (0.059)&0.801 (0.61, 1.12)&0.001&0.059\\\cline{3-12}
&&\begin{tabular}{c}Weak\\NCO\end{tabular}&{\bf Proposed}&0.845 (0.346)&0.825 (-5.42, 5.22)&0.045&0.349&0.806 (0.125)&0.810 (-0.57, 1.54)&0.006&0.126\\\hline\hline
\begin{tabular}{c}Specify\\valid IVs\end{tabular}&\begin{tabular}{c}Strong\\IV\end{tabular}&-&{\bf Liao, 2013}&0.801 (0.087)&0.802 (0.49, 1.34)&0.001&0.087&0.829 (0.107)&0.803 (0.66, 1.28)&0.029&0.111\\\cline{4-12}
&&&\begin{tabular}{c}{\bf DiTraglia},\\{\bf 2016}\end{tabular}&0.799 (0.081)&0.801 (0.50, 1.04)&0.001&0.081&0.799 (0.040)&0.799 (0.66, 0.92)&0.001&0.040\\\cline{4-12}
&&&{\bf GMM}&0.799 (0.081)&0.801 (0.50, 1.04)&0.001&0.081&0.799 (0.040)&0.799 (0.66, 0.92)&0.001&0.040\\\cline{2-12}
&\begin{tabular}{c}Weak\\IV\end{tabular}&-&{\bf Liao, 2013}&0.823 (0.186)&0.815 (0.35, 2.10)&0.023&0.188&0.879 (0.252)&0.811 (0.61, 1.99)&0.079&0.265\\\cline{4-12}
&&&\begin{tabular}{c}{\bf DiTraglia},\\{\bf 2016}\end{tabular}&0.806 (0.139)&0.811 (0.35, 1.15)&0.006&0.139&0.803 (0.070)&0.804 (0.61, 1.00)&0.003&0.070\\\cline{4-12}
&&&{\bf GMM}&0.806 (0.139)&0.811 (0.35, 1.15)&0.006&0.139&0.803 (0.070)&0.804 (0.61, 1.00)&0.003&0.070\\\hline\hline
\begin{tabular}{c}Unspecify\\valid IVs\end{tabular}&\begin{tabular}{c}Strong\\IV\end{tabular}&-&{\bf Liao, 2013}&1.359 (0.214)&1.368 (0.76, 1.94)&0.559&0.599&1.270 (0.156)&1.259 (0.93, 1.67)&0.470&0.496\\\cline{4-12}
&&&\begin{tabular}{c}{\bf DiTraglia},\\{\bf 2016}\end{tabular}&1.200 (0.143)&1.208 (0.72, 1.64)&0.400&0.425&1.215 (0.064)&1.213 (1.01, 1.42)&0.415&0.420\\\cline{4-12}
&&&{\bf GMM}&1.215 (0.125)&1.213 (0.87, 1.64)&0.415&0.434&1.215 (0.064)&1.213 (1.01, 1.42)&0.415&0.420\\\cline{2-12}
&\begin{tabular}{c}Weak\\IV\end{tabular}&-&{\bf Liao, 2013}&2.844 (0.588)&2.780 (0.98, 7.93)&2.044&2.127&2.647 (0.533)&2.796 (1.32, 4.03)&1.847&1.923\\\cline{4-12}
&&&\begin{tabular}{c}{\bf DiTraglia},\\{\bf 2016}\end{tabular}&2.715 (0.538)&2.632 (1.17, 8.08)&1.915&1.989&2.703 (0.223)&2.687 (2.10, 3.77)&1.903&1.916\\\cline{4-12}
&&&{\bf GMM}&2.715 (0.538)&2.632 (1.37, 8.08)&1.915&1.989&2.703 (0.223)&2.687 (2.10, 3.77)&1.903&1.916\\\hline
\end{tabular}
}
\label{tab21}
\end{center}
\end{table}
\end{landscape}
\begin{landscape}

\begin{figure}[h]
\begin{center}
\begin{tabular}{c}
\includegraphics[width=24cm]{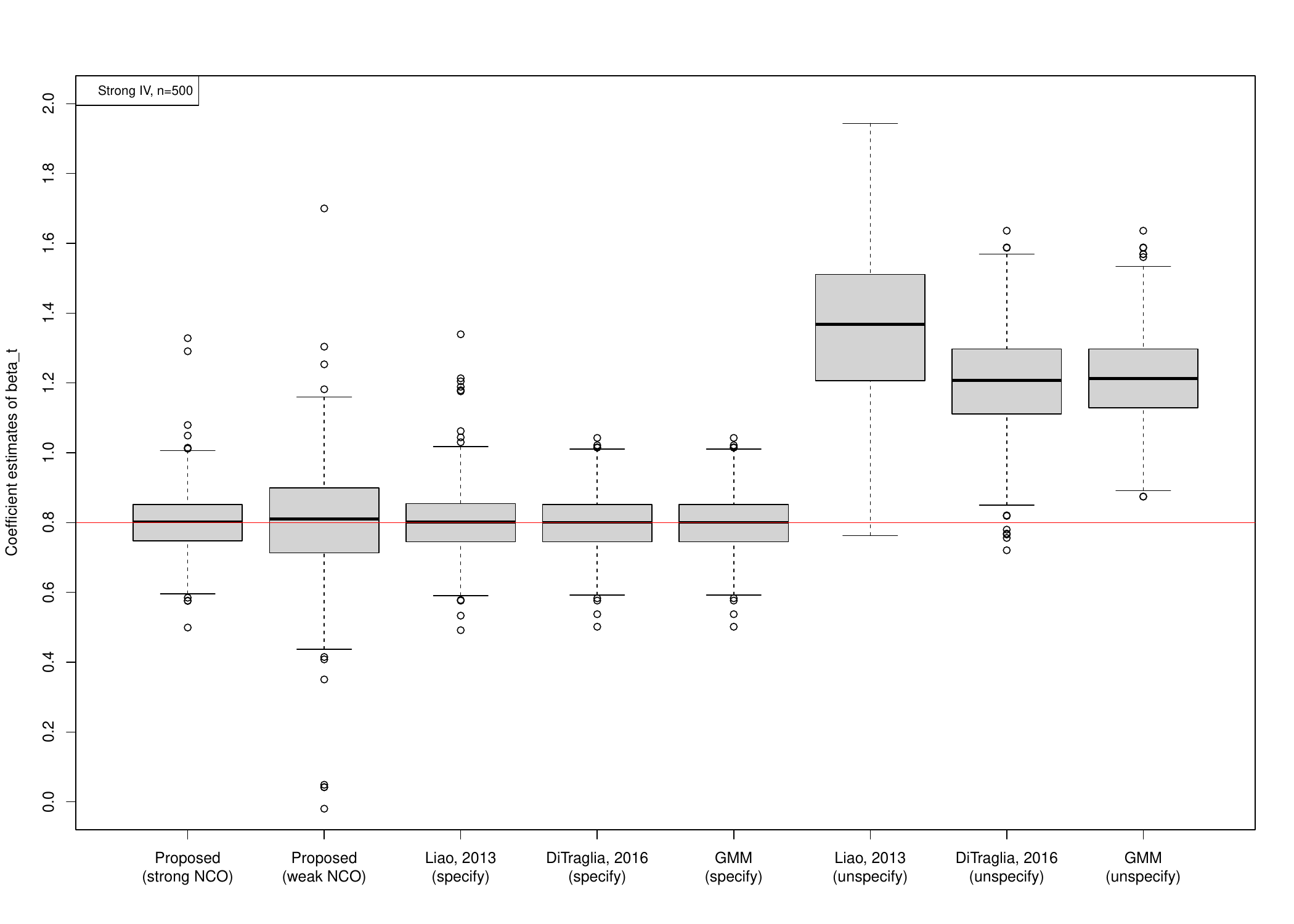}
\end{tabular}
\caption{Boxplot of estimators for causal effects (strong IV, $n=500$)}
\label{fig1}
\end{center}
\end{figure}

\begin{figure}[h]
\begin{center}
\begin{tabular}{c}
\includegraphics[width=24cm]{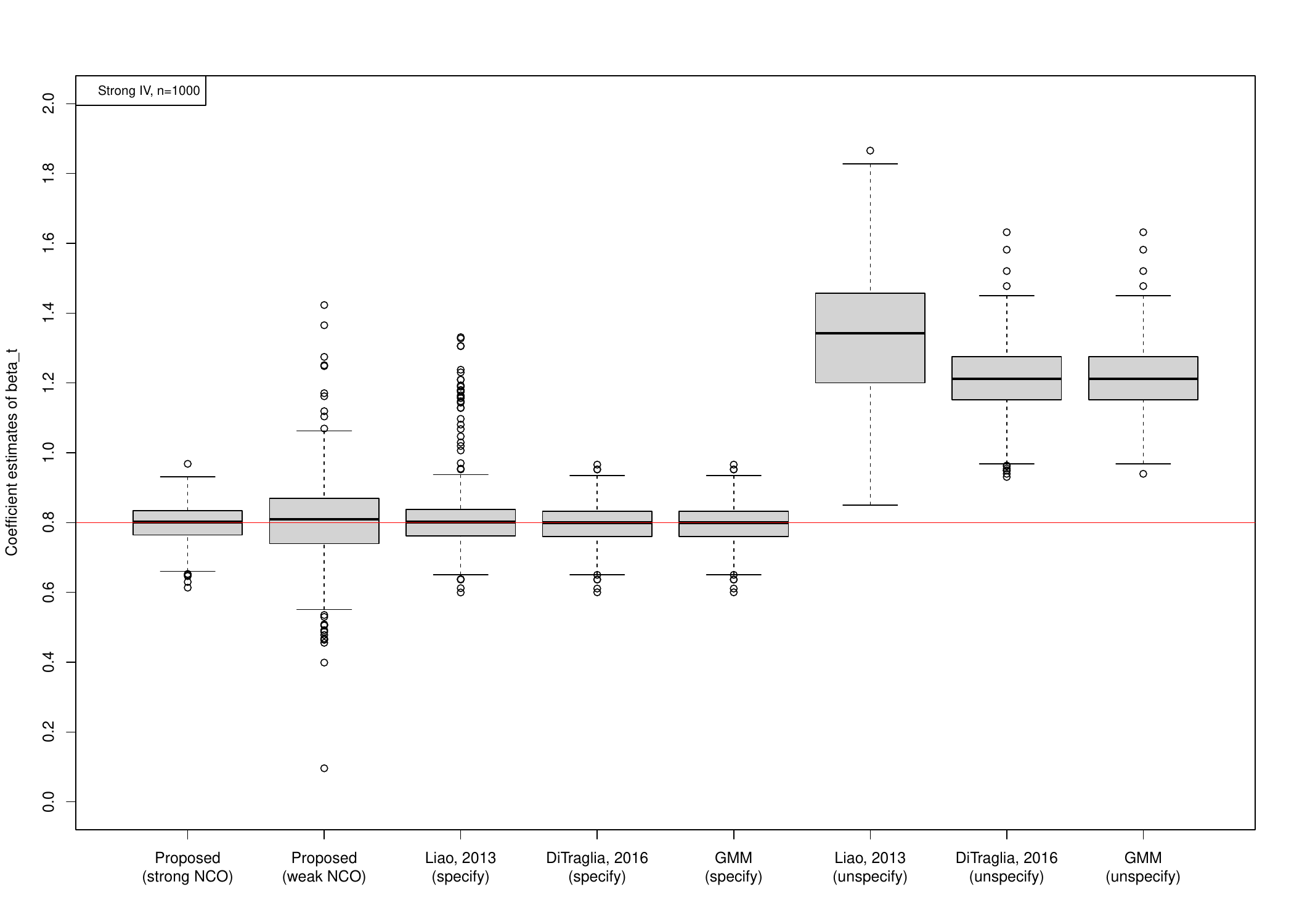}
\end{tabular}
\caption{Boxplot of estimators for causal effects (strong IV, $n=2000$)}
\label{fig2}
\end{center}
\end{figure}

\begin{figure}[h]
\begin{center}
\begin{tabular}{c}
\includegraphics[width=24cm]{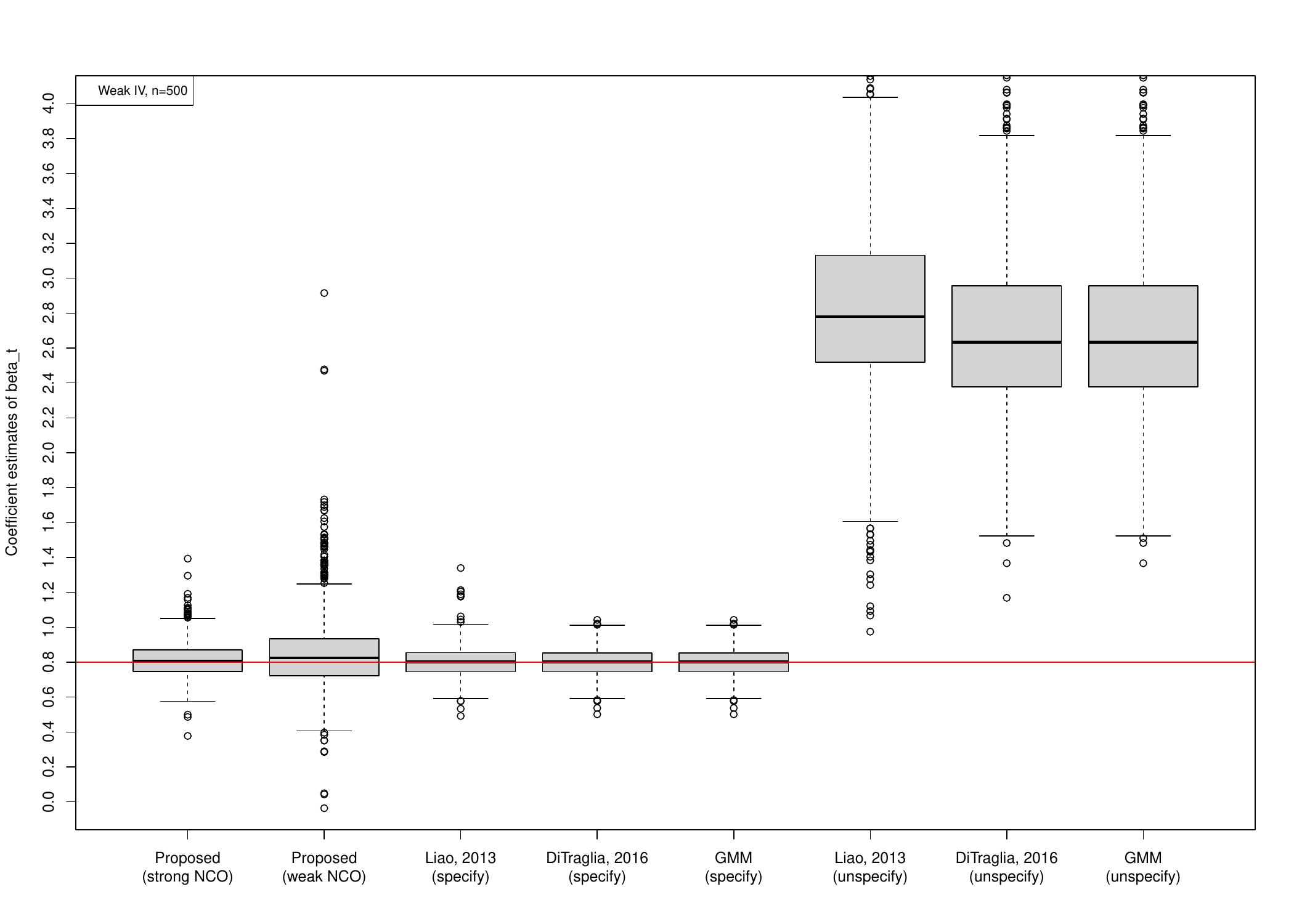}
\end{tabular}
\caption{Boxplot of estimators for causal effects (weak IV, $n=500$)}
\label{fig3}
\end{center}
\end{figure}

\begin{figure}[h]
\begin{center}
\begin{tabular}{c}
\includegraphics[width=24cm]{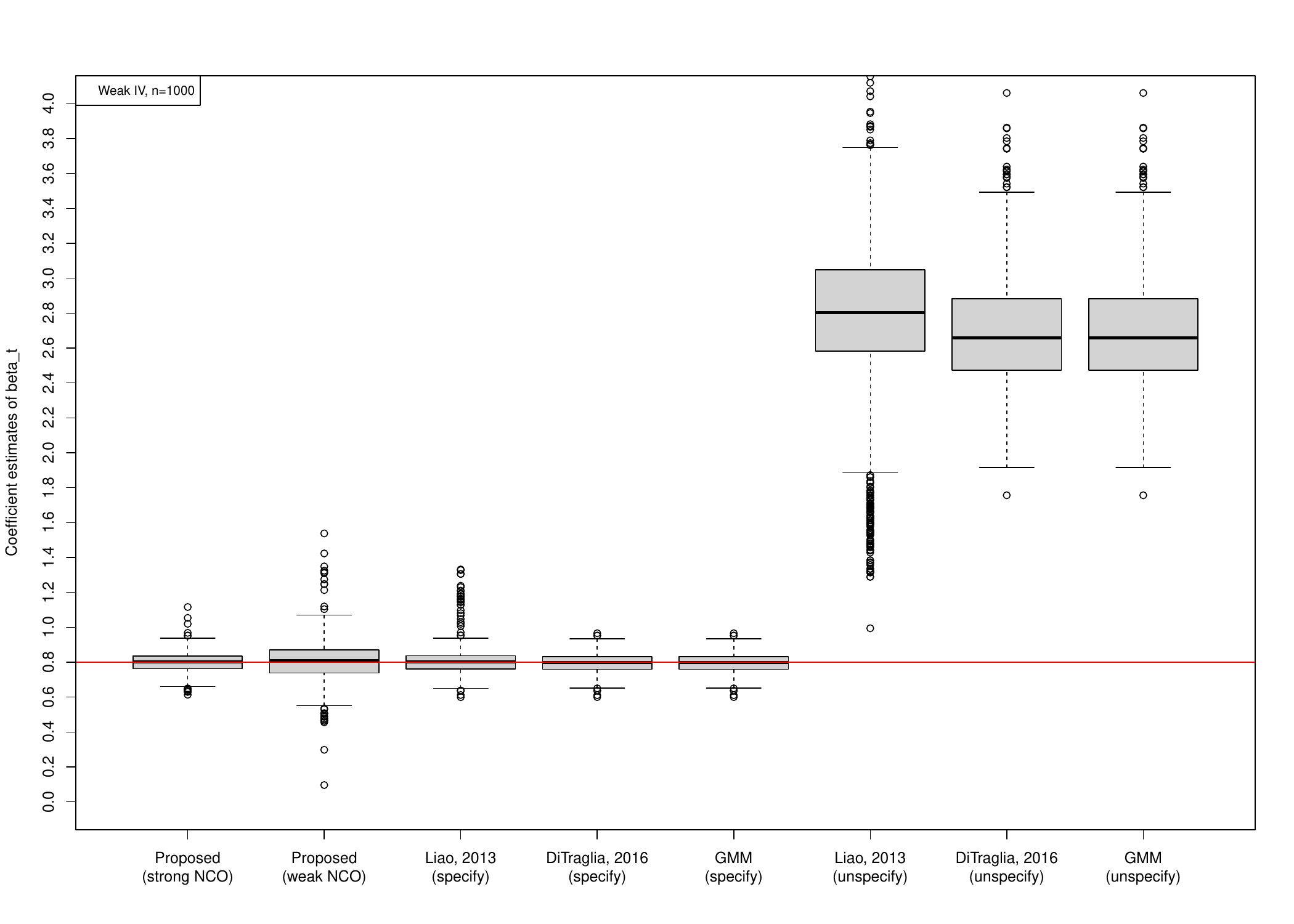}
\end{tabular}
\caption{Boxplot of estimators for causal effects (weak IV, $n=2000$)}
\label{fig4}
\end{center}
\end{figure}
\end{landscape}

\subsection{Figures of mendelian randomization setting}

\begin{landscape}
\begin{figure}[h]
\begin{center}
\begin{tabular}{c}
\includegraphics[width=20cm]{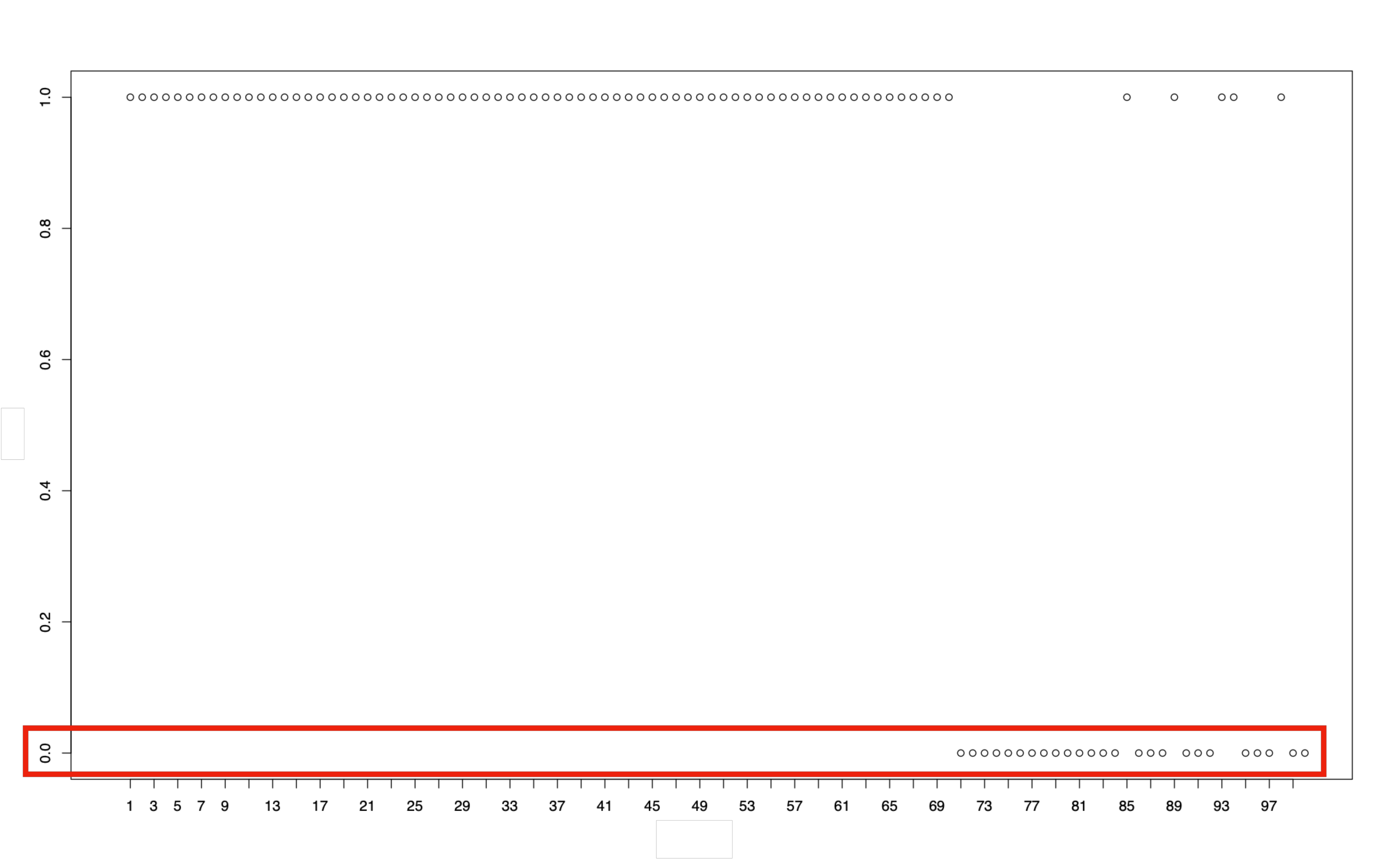}
\end{tabular}
\caption{One example of selected valid IVs for our proposed method (Valid IVs are majority situation)}
\label{fig5}
\end{center}
\begin{description}
\item{\bf Note} Red box is not used IVs
\end{description}
\end{figure}

\begin{figure}[h]
\begin{center}
\begin{tabular}{c}
\includegraphics[width=20cm]{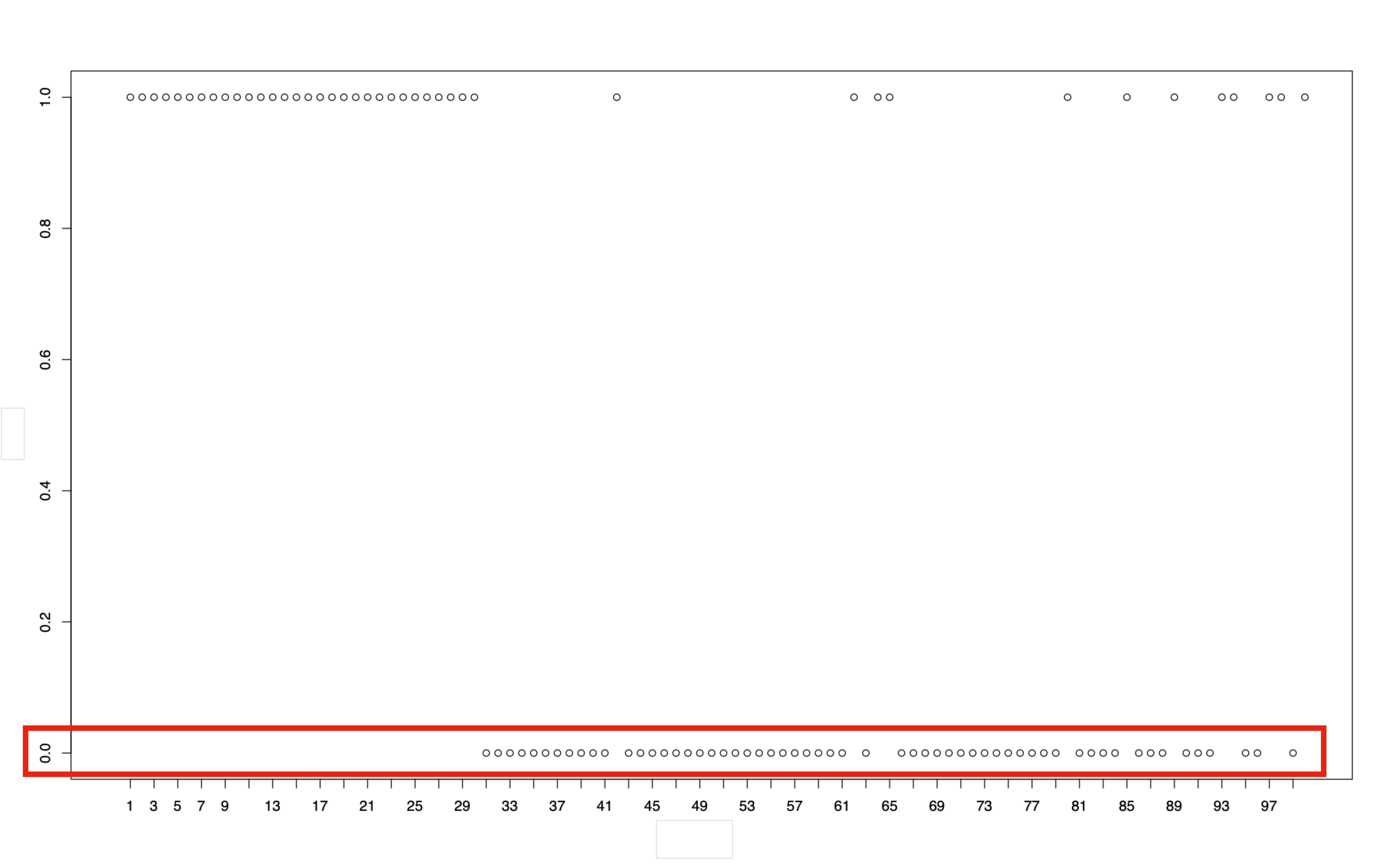}
\end{tabular}
\caption{One example of selected valid IVs for our proposed method (Invalid IVs are majority situation)}
\label{fig6}
\end{center}
\begin{description}
\item{\bf Note} Red box is not used IVs
\end{description}\end{figure}
\end{landscape}

\end{document}